\documentclass[screen, manuscript]{acmart}  
\usepackage{amsfonts,amsmath,tikz,textcomp}
\usepackage{algorithm,algpseudocode,algorithmicx}
\usepackage{wrapfig} 
\usetikzlibrary{arrows,automata}
\usepackage[english]{babel}
\usepackage{hyperref}
\usepackage[caption=false]{subfig}
\usepackage{multirow, adjustbox}

\usepackage{blindtext}
\usepackage{xpatch}
\makeatletter
\xpatchcmd{\ps@firstpagestyle}{Manuscript submitted to ACM}{}{\typeout{First patch succeeded}}{\typeout{first patch failed}}
\xpatchcmd{\ps@standardpagestyle}{Manuscript submitted to ACM}{}{\typeout{Second patch succeeded}}{\typeout{Second patch failed}}    \@ACM@manuscriptfalse
\makeatother

\renewcommand\footnotetextcopyrightpermission[1]{} 
\setcopyright{none}
\definecolor{upppurple}{RGB}{140, 56, 99}
\definecolor{uppinv}{RGB}{167, 66, 168}
\definecolor{uppmagenta}{RGB}{231, 66, 128}
\definecolor{uppgreen}{RGB}{66, 168, 72}
\definecolor{uppsynchr}{RGB}{66, 160, 168}
\definecolor{uppupdate}{RGB}{66, 66, 168}

\usepackage{xpatch}

\makeatletter
\xpatchcmd{\ps@firstpagestyle}{Manuscript submitted to ACM}{}{\typeout{First patch succeeded}}{\typeout{first patch failed}}
\xpatchcmd{\ps@standardpagestyle}{Manuscript submitted to ACM}{}{\typeout{Second patch succeeded}}{\typeout{Second patch failed}}    \@ACM@manuscriptfalse
\makeatother

\newcommand{\clbox}[1]{
\begin{tikzpicture}				
		\tikzstyle{every state}=[rectangle, fill=#1,draw=black,thick,scale=0.2]
		\node[state] at (0,0) (s0) {};
	\end{tikzpicture}
}
\newcommand{\ccounter}{c}
\newcommand{\numobs}{N}
\newcommand{\perr}[2]{
\def\arraystretch{0.35}
$\begin{array}{c} \\ #1 \\ \scalebox{0.6}{$\pm #2$} \end{array}$
\def\arraystretch{1}
}
\newcommand{\perrthin}[2]{
\def\arraystretch{0.35}
\hspace{-0.18cm}
$\begin{array}{c} \\ #1 \\ \scalebox{0.6}{$\pm #2$} \end{array}$
\hspace{-0.18cm}
\def\arraystretch{1}
}

\begin{document}

\title{A spatial algorithm for the analysis of transportation systems using statistical model checking}
\author{Dani\"el Reijsbergen}
\affiliation{Singapore University of Technology and Design}
\email{daniel_reijsbergen@sutd.edu.sg}

\author{Stephen Gilmore}
\affiliation{University of Edinburgh}
\email{stg@inf.ed.ac.uk}

\authorsaddresses{}

\begin{abstract}

We present an automated methodology for using Automatic Vehicle Location measurements of public transportation vehicles to construct a probabilistic model. The model not only allows for accurate evaluation of service performance, but also makes it possible to study the effects of system modifications \emph{a priori}. The methodology is almost entirely agnostic to otherwise important details of the service --- in particular its route and the location of stops. Instead, it infers this from the data using automated map generation techniques. The behaviour of vehicles in the model is analysed using computer simulation combined with statistical model checking. We present two case studies involving the Airlink service in Edinburgh and the Bellevue Express in Seattle. To demonstrate the usefulness of the approach, we analyse the impact of the scheduling strategies of bus holding and speed modification on the Airlink's performance. The data and code used to create the figures are publicly available online.

\end{abstract}

\maketitle


\section{Introduction} \label{sec: introduction}

The wealth of data produced by the increasing digitisation of urban transportation systems has made it considerably easier to evaluate their performance. Modern GPS and communication technologies allow for a level of monitoring of system components --- particularly vehicles --- that was previously unimaginable. At the same time, improvements in formal verification techniques have enabled increasingly realistic system models to be analysed in a rigorous manner. There is a synergy between these two developments in the sense that the ability to handle more complex models is largely futile if there is insufficient data for realistic parameterisation. Hence, there is a need for a methodology that can translate the vast datasets generated by `smart' transportation systems into properly parameterised formal models, which can then be analysed by modern analysis tools. 

In this paper, we present a novel technique that uses Automatic Vehicle Location (AVL) data to build and parameterise a formal model for vehicle movements along a route. This model can be used to determine whether a service satisfies a given performance requirement, \emph{e.g.}, one set by regulators. Additionally, by making alterations to the model, planners can study the impact of changes to the network on its performance \emph{a priori}. Property evaluation is done using \emph{model checking} \cite{clarke1986automatic,clarke1999model,baier2008principles}. Model checking is a formal verification technique that essentially consists of two steps: (1) specification of a formal {model} representing the system and (2) evaluation of whether the model satisfies a formally specified \emph{property}, expressed as a formula in a suitable logic.
Since transportation systems are strongly subject to random effects (\emph{e.g.}, traffic conditions or passenger numbers at bus stops), we use \emph{stochastic} models. Our models are spatial and \emph{patch-based}: the service under consideration is partitioned into a discrete set of patches, and vehicles move from one patch to the next such that the time spent in each patch is modelled using a formally defined probability distribution. 
Our methodology is almost entirely agnostic to route information, as the AVL data is used both to determine the patch structure and to parameterise the probability distributions of the times spent within them.
 The properties of interest are expressed using the language MultiQuaTeX, which allows us to use the powerful and versatile statistical model-checking algorithms of the MultiVeStA tool \cite{sebastio2013multivesta}.

We illustrate our methodology using case studies that focus on performance measures for \emph{frequent} services, defined as services for which more than six bus departures are timetabled each hour. Unlike infrequent services, for which punctuality (\emph{i.e.}, timetable adherence) is the main performance criterion, the main metrics for frequent services involve the regularity of the \emph{headways}, \emph{i.e.}, the time between subsequent vehicle arrivals at a stop. We focus on the common notion of \emph{Excess Waiting Time} (EWT), but to demonstrate the broad applicability of our method we also consider two measures specified by the Scottish government \cite{bpips}. We perform \emph{two} case studies involving frequent services, one in Edinburgh (Scotland, UK) and one in Seattle (Washington, USA). In addition to evaluating the current performance of the services, we study the impact of performance improvement strategies --- \emph{bus holding} and \emph{speed modification}. Both strategies depend on parameters set by the operator, and these greatly determine the strategy's success \cite{moreira2015improving}. The ability to discover  a good choice of control parameters is a powerful contribution of our framework. Both the code, which is open source, and the data used to conduct the experiments presented in this paper are available online via \url{http://dx.doi.org/10.7488/ds/1472} and \url{http://dx.doi.org/10.7488/ds/1470} respectively.

The use of our approach has following key advantages over other techniques known in the literature:
\begin{itemize}
	\item The numerical efficiency of our approach allows for quick evaluation of the service requirements. This is particularly helpful for operators who seek to optimise timetables or control measures using a wide-ranging parameter sweep.
	\item The methodology is almost completely \emph{agnostic} to information about the route, and learns it from the data instead. This makes the methodology applicable even when this information is outdated, wrong (\emph{e.g.}, if due to road works buses take a different route than prescribed), or unavailable.
	\item The spatial nature of the patch identification step makes it easier for planners to make spatially-informed changes to the model (\emph{e.g.}, the imposition of speed limits in specific areas --- see also \cite{reijsbergen2015probabilistic}), or bus control measures that can only be imposed in certain areas (\emph{e.g.}, lowering bus speeds only in areas that are not heavily congested). 
	\item Informative maps that identify regions of different vehicle movement behaviour are created as a by-product of the parameter fitting procedure.
	\item Our methodology can be used to evaluate \emph{any} performance property that can be expressed in the highly expressive formal language MultiQuaTeX. Although we focus on EWT and the two measures used by the Scottish government, extensions to other examples, \emph{e,g.}, on-time performance for infrequent services, or coefficients of variation of headway ratios \cite{moreira2016online}, are straightforward.
\end{itemize}
In particular, the patch identification technique and the use of model checking to evaluate performance properties are the main contributions of this paper. As part of the case studies, we present a detailed analysis of the appropriate stochastic model for the time spent by vehicles in each of the patches. We find that if the data is sufficiently regular (\emph{e.g.}, in patches outside the city centre and during non-peak periods), simple Erlang distributions give a good fit, but that in other cases we need distributions with a heavier tail (\emph{e.g.}, hyper-Erlang).

The structure of this paper is as follows. In Section~\ref{sec: model setting} we give an overview of the datasets and the related scientific literature. In Section~\ref{sec: preliminaries} we present the theoretical preliminaries, in particular a discussion of phase-type modelling and the specification of performance properties. In Section~\ref{sec: algorithm}, we present the main algorithm of the paper. In Section~\ref{sec: results} we provide the numerical results, and Section~\ref{sec: conclusions} concludes the paper.


\section{Background \& Related Work} \label{sec: model setting}

Since our approach is data-driven, we start with a discussion of the two datasets (Edinburgh and Seattle) in Sections~\ref{sec: data edinburgh}~and~\ref{sec: data seattle}. We provide an overview of related literature in Section~\ref{sec: related work}.

\subsection{Edinburgh Dataset} \label{sec: data edinburgh}

The main dataset used in this paper was collected and provided to us by the Lothian Buses company, which operates an extensive bus network in Edinburgh. The provided dataset consists of bus GPS measurements obtained for the full fleet of 745 buses between 28th January 2014 at 11:31:14 and 30th January 2014 at 12:38:31. 
The measurements were collected using the AVL system built into each bus for the purpose of monitoring the buses and providing live arrival time predictions at bus stops. The AVL technology reports to a Real Time Passenger Information (RTPI) system called Bustracker, operated by the French company Cofely Ineo. The measurements are collected through centralised pull requests made by a central server in France, with measurements coming in roughly every 35 seconds. This type of data can be publicly accessed live via the MyBustracker API.\footnote{\url{http://www.mybustracker.co.uk/}} 

The bus GPS measurements in the dataset have already gone through a data pre-processing step. As a result, the measurements are very accurate, although some quirks can be observed, \emph{e.g.}, buses `flying' from one point in the city to another (due to interpolation), or buses seemingly going over 50mph on Princes Street, a busy shopping street in the city centre. The measurements contain bus vehicle identifiers (fleet numbers). However, since buses can serve different routes on different days (sometimes even on a single day if the vehicle has been assigned to a different service during peak hours), we do not always know for a specific measurement which route the bus is serving at the moment that the measurement was taken. 
In this paper, we avoid this complication by focusing on a specific route in Edinburgh, namely the Airlink service which connects the city centre to the airport. Since the buses on this route are coated in a distinct livery, they cannot be assigned to other routes. For the Airlink service, there are unfortunately two large GPS ``shadows'' (\emph{i.e.}, areas where the GPS antennas have limited connectivity) in which the buses are not observed. This has consequences for our algorithm, as we discuss in more detail in Section~\ref{sec: map generation}. To cover the gaps, we perform linear interpolations between measurements, which is possible because we have measurement timestamps (additionally, the measurements are chronologically sorted in the dataset).\footnote{Note that since buses are never observed in the gaps, we cannot interpolate using averaged crossing behaviour from earlier observations as is done in, \emph{e.g.}, \cite{mendes2015validating}.} Since the average times needed by buses to clear route segments depend on the time of day (\emph{e.g.}, crossing the city centre takes much more time during rush hours), we focus on a specific period of the day, namely between 10AM and 3PM. Although it is possible to incorporate time-dependent average speeds in our model (using time-inhomogeneous Markov chains \cite{stroock2013introduction}), this is left to future work.

In principle, our approach can be applied to any GPS dataset that is in the format of the one provided, \emph{i.e.}, consisting of four data fields (bus ID, latitude, longitude, time) per measurement. 
The spatial data in the Edinburgh dataset is not given by default in longitude/latitude, but in the Eastings/Northings coordinates used for the Ordnance Survey National Grid in the UK. They can be converted into longitude/latitude using the Jcoord library in Java.\footnote{\url{http://www.jstott.me.uk/jcoord/}} The date/time format needs to be converted to UNIX timestamps --- a custom format can be set in the supplementary programming code (specifically in the AVLDataProcesser class).

\subsection{Seattle Dataset} \label{sec: data seattle}

In addition to the Edinburgh dataset, we also use a publicly available dataset involving buses in Seattle \cite{rice2003data}. This dataset spans roughly one month of data (November 2001). One advantage of the Seattle dataset is that each measurement includes both a vehicle and service identifier, meaning that it is trivial to isolate routes. However, the Seattle dataset is considerably noisier than the Edinburgh dataset, as we discuss in more detail in Section~\ref{sec: seattle}. Although entries in the Seattle dataset are chronologically sorted per bus, the dataset as a whole is not, meaning that for a pair of successive measurements the earlier measurement does not necessarily appear first if the measurements involve different buses. The location measurements are given in a bespoke $x,y$-coordinate system that does not trivially map to latitude/longitude coordinates. To aid visual inspection of the data, our code has support for approximate translation between the two coordinate systems, but the algorithm of Section~\ref{sec: algorithm} is shift- and scale-invariant to the measurements --- \emph{i.e.}, if an offset or zoom is applied to all measurements simultaneously, then the results from our algorithm remain the same.

We focus on the Bellevue Express (Route 550), which is operated by SoundTransit and connects downtown Seattle to the Bellevue area in the east across Lake Washington. We have chosen the Bellevue Express for two reasons. First, the Bellevue express is a frequent service during the evening rush hour period between 4PM and 6PM on working days. Second, the heterogeneity of the route --- two densely populated areas connected by a long stretch of highway --- makes it very suitable to demonstrate the differences between the patches, and the traffic conditions inside the densely populated areas during the rush hour period make the parameter fitting within the patches more difficult than for the more predictable Airlink service.

\subsection{Related Work} \label{sec: related work}

The use of AVL data to evaluate or improve public transport performance is an active research field: see \cite{moreira2015improving} for a recent literature overview. In particular, a number of probabilistic models have been proposed for the purpose of vehicle travel time prediction. Examples include models based on ARIMA time series \cite{suwardo2010arima}, Kalman filters \cite{shalaby2003bus}, and basic Markov models \cite{lin2004modeling}. In particular, the latter model represents the delay at each stop in the system using a Markov chain with a state for each minute of delay. Models to predict travel times can be used to infer models for vehicle headways at stops, which allows for the calculation of headway-based performance metrics. The model of this paper explicitly uses spatial information --- hence it is considerably easier in our setting to make spatially-informed alterations to the system, such as the introduction of a speed limit or space-based bus strategies. Furthermore, our approach is agnostic to stop data.

Other examples of the use of AVL data to evaluate system performance include \cite{cevallos2011using} and \cite{cevallos2012using}, in which Gaussian distributions are used to evaluate and optimise timetable performance of a bus system in Miami. In \cite{tetreault2010estimating}, an approach was presented for minimising service run times as a function of stop locations, where the run times were computed using AVL and Automatic Passenger Counting (APC) data from Montr\'eal, Canada. See also \cite{hawas2013simulation} for an approach to evaluate travel times and passenger numbers using a micro-level simulation which takes into account, for example, lanes and intersections. In \cite{moreira2016online}, AVL data is combined with a machine learning procedure to predict travel times and headways. This is then used to predict bus bunching --- \emph{i.e.}, the phenomenon where a delayed bus keeps accruing more delay due to the greater number of passengers at subsequent stops, to the point where it is in very close proximity to the next bus serving the same route. Such a prediction algorithm for bus bunching can inform a control strategy. It would be interesting to see whether these strategies can be modelled using a formalism that is compatible with model-checking tools, so that service requirements involving a broad variety of performance measures can be checked efficiently using high-performance evaluation algorithms.

The current approach of dividing a bus route into patches and fitting Erlang or more general phase-type distributions, \emph{e.g.}, hyper-Erlang to the time spent by buses in the patches can also be found in \cite{reijsbergen2015patch} and \cite{vissat2015finding}. 
A formal discussion of the performance metrics for frequent services used in this paper is given in \cite{reijsbergen2014formal}. Here, a time series model was fitted to the bus arrivals, meaning that the movement of buses serving the route became lost in the abstraction. Finally, in  \cite{reijsbergen2015probabilistic} different probability distributions for the time spent in patches were compared, and the resulting model was used to evaluate the impact of a planned speed limit reduction in Edinburgh. One of the aims of this paper is to unify the methodology developed over the course of these papers.

The patch identification technique presented in this paper incorporates automated map generation techniques. For a recent overview of these techniques, see \cite{ahmed2015comparison}. In particular, we make heavy use of the work of Biagioni \emph{et al.}\ on the EasyTracker project \cite{biagioni2012map, biagioni2011easytracker}.  The method proposed in \cite{biagioni2011easytracker} is able to perform route map generation, stop location identification, timetable construction and arrival time prediction at stops using only AVL data. This paper expands on their approach by using it to inform the construction of a simulation and performance evaluation model.

In general, the question of how best to model the time spent by vehicles on route segments has been studied for many decades \cite{berry1951distribution}, and is still being actively researched \cite{skabardonis2005real,hofleitner2012probability,cao2014modeling}. Among these papers, \cite{hofleitner2012probability} and \cite{cao2014modeling} are particularly relevant as the multimodal distributions reported in those papers resemble some of the distributions encountered in our case studies (\emph{e.g.}, one can compare our Figure~\ref{fig: gof investigation} to Figure 6 of \cite{hofleitner2012probability}, or to Figure~5 of \cite{cao2014modeling}). Although the modelling choices in those papers (\emph{e.g.}, the use of truncated distributions of \cite{cao2014modeling}) are interesting, the use of the Erlang and hyper-Erlang distribution gives a sufficiently good fit in our setting, as we discuss further in Section~\ref{sec: results}.


\section{Preliminaries} \label{sec: preliminaries}

In this section, we briefly discuss the formal principles needed in the later sections. We first discuss the basic stochastic modelling framework that we will use for the patch crossing times in Section~\ref{sec: patch-based models}. In Section~\ref{sec: model checking basics}, we discuss the model checking techniques that we will use to evaluate performance properties.

\subsection{Patch-Based Vehicle Movement Models} \label{sec: patch-based models}

In our paper, we represent the time spent by vehicles inside the patches using a \emph{phase-type} distributions, which have a long history \cite{altiok1985phase,asmussen1996fitting} of being used to succinctly represent general distributions in a wide range of applications (\emph{e.g.}, \cite{faddy1999analysing,bladt2005review,neuts1981use}). 
In our context, this means that the process of a bus moving across a patch can be modelled using a sequence of \emph{phases} such that the time to complete each phase is \emph{exponentially distributed}. The time $T$ spent in a single phase is exponentially distributed with \emph{rate} $\lambda>0$ if its \emph{Probability Density Function} (PDF) is given by
\begin{equation}
f_T(t;\lambda) = \lambda e^{-\lambda t} \label{eq: exp pdf}
\end{equation}
for $t \geq 0$ and by $0$ otherwise. The expected amount of time spent in a phase is given by $\frac{1}{\lambda}$ --- hence, large values of the rate $\lambda$ mean that the time spent in a phase is small on average. The PDF $f_T(t)$ of a random variable $T$ determines the probability of observing values from a small interval around $t$, and can also be used to measure how well a fitted probability distribution corresponds to the data.  In the appendix, we use the related function $F_T(t) = \int_{0}^{t} f_T(\tau) d\tau$, called the \emph{Cumulative Density Function} (CDF) to compare fitted to empirical distributions.

The exponential distribution has several characteristics that often make it unsuitable for modelling patch crossing times. It has a high standard deviation relative to its mean (called the \emph{coefficient of variation}), its mode at 0, and it is \emph{memoryless}, meaning that the amount of time already spent in a patch gives no information on the probability distribution of the remaining time.  A straightforward generalisation, called the \emph{Erlang distribution} is often more appropriate. Given a sequence $T_1,\ldots, T_k$ of exponentially distributed random variables with rate $\lambda>0$, \mbox{$S = T_1 + \ldots + T_k$} is Erlang-distributed with rate $\lambda$ and shape $k \in \mathbb{N}$. Its PDF is given by
\begin{equation}
f_S(s;\lambda,k) = \frac{\lambda^k s^{k-1} e^{-\lambda s}}{(k-1)!} \label{eq: erlang pdf}
\end{equation}
for $s \geq 0$ and $0$ otherwise. In this paper, we focus primarily on the basic Erlang distribution, both for simplicity and because the fitting results are already good as witnessed by Figures~\ref{fig: cdf plots}~and~\ref{fig: cdf plots seattle}. However, several generalisations of the Erlang distribution exist that can further improve the goodness-of-fit to observation data, at the cost of added model complexity. In \cite{reijsbergen2015probabilistic}, two extensions were considered: the hyper-Erlang distribution, which allows for the use of the tool HyperStar \cite{reinecke2012hyperstar}, and a shifted Erlang distribution that is close to the distribution recommended by the Traffic Engineering Handbook \cite{pline1992traffic}. In particular, the hyper-Erlang distribution is a \emph{mixture} of two or more Erlang branches. The PDF of a random variable $Z$ with an $m$-branch hyper-Erlang distribution with rates $\vec{\lambda} = (\lambda_1,\ldots,\lambda_m)$, shapes $\vec{k} = (k_1,\ldots,k_m)$, and $\vec{\alpha} = (\alpha_1,\ldots,\alpha_m)$, is given by
\begin{equation}
f_Z(z;\vec{\lambda},\vec{k},\vec{\alpha}) = \sum_{i=1}^m \alpha_i \frac{\lambda_i^{k_i} z_i^{k_i-1} e^{-\lambda_i z}}{(k_i-1)!} \label{eq: erlang pdf}
\end{equation}
for $z \geq 0$ and $0$ otherwise (here, $\lambda_1>0$, $m_i\in\mathbb{N}$, and $\alpha_i > 0$ for all $i \in \{1,\ldots,m\}$, and $\sum_{i=1}^m \alpha_i = 1$).
The software implementation used to conduct the experiments of Section~\ref{sec: results} is sufficiently modular to add other distributions (\emph{e.g.}, lognormal, the distributions from \cite{hofleitner2012probability} and \cite{cao2014modeling}) when required. 

In general, any probabilistic model described by a system state transitioning between phases (possibly skipping phases or going back) such that the time spent in each phase is exponentially distributed is called a \emph{Continuous-Time Markov chain (CTMC)}. The probability distribution of the time until the system reaches a specific phase in the CTMC is called a phase-type distribution, of which the exponential and Erlang distributions are rudimentary examples.

\subsection{Model Checking} \label{sec: model checking basics}

Having specified the model, properties of interest can be expressed using a formal specification language. The choice of specification language depends on the modelling formalism used. For example, for CTMCs the most commonly used property specification language is Continuous Stochastic Logic (CSL) \cite{aziz2000model, baier2003model}. 
In this paper, we use the more general language MultiQuaTEx \cite{sen2005vesta, sebastio2013multivesta},  which we extend with a notion of steady-state properties. We do not aim to describe MultiQuaTEx in full, but only those language features used in Section~\ref{sec: model checking}. In the following, we will assume that we are given a model simulator~$\texttt{s}$, which can be queried using statements of the form $\texttt{s.rval("Y")}$, where $Y$ is either a location label (in which case the returned value is $1$ if the vehicle is in that location, otherwise $0$), or the name of a clock or integer counter (in which case the value of the clock or counter is returned).  Furthermore, specification of functions and if-then-else statements is allowed. For example, the query
\[
  \texttt{isYAboveThreshold() = if \{s.rval("Y") > 5 \} then 1 else 0 fi;}
\]
returns $1$ if the value returned by the query \texttt{s.rval("Y")} is above the threshold value $5$ and $0$ otherwise. As a special case, \texttt{s.rval("time")} returns the value of the global system clock.
For model checking, we are interested in assertions of the form
\begin{equation}
  \texttt{ S [ F(), "C"] < p;} \label{eq: steady-state quatex}
\end{equation}
In words, this asserts that the steady-state value of the random variable defined through the function \texttt{F()}, with time given by the clock \texttt{C}, is smaller than a given value \texttt{p} between 0 and 1.  To make this formal, let ${x}(t)$ represent the global state (recall that this is, for each automaton, its current location, combined with the values of all local clocks and counters) as a function of the global clock $t$. Then \eqref{eq: steady-state quatex} asserts that the value
\begin{equation}
	\lim_{T \rightarrow \infty} \frac{1}{T} \int_{0}^{T} F(x(t)) dC(t) \label{eq: steady-state fC}
\end{equation}
is smaller than $p$. The evolution of our system can be represented by a discrete-event process $(t_0, x_0), (t_1, x_1), \ldots, (t_N, x_N)$ such that the global state $x(t)$ equals $x_{i}$ for all $t \in [t_i, t_{i+1})$, \eqref{eq: steady-state fC} can be rewritten to 
\begin{equation*}
   \frac{1}{t_N - t_0} \sum_{i=1}^{t_N} (C(t_i) - C(t_{i-1})) F(x_{i-1})
\end{equation*}
where $t_0 = 0$ and $t_N = T$. The latter equation is what we have implemented in the simulation engine BusSimulator discussed in Section~\ref{sec: algorithm}.

Assertions written in the form of \eqref{eq: steady-state quatex} can be model-checked using \emph{statistical model checking} \cite{younes2002probabilistic}, which uses computer simulation to produce a statistically justified statement about whether the properties hold. Exact numerical computation of steady-state probabilities as mentioned above is often infeasible, so we use a purpose-built simulator combined with the model-checking tool MultiVeStA \cite{sebastio2013multivesta} (and in particular its steady-state model checking functionality \cite{gilmore2017transient}) to obtain the results of Section~\ref{sec: results}. In particular, the steady-state properties are evaluated using the batch means method \cite{alexopoulos1996implementing}, as implemented in the ASAP3 algorithm \cite{steiger2005asap3}.


\section{Main Algorithm} \label{sec: algorithm}

\newcommand{\intp}{\delta}
\newcommand{\gauss}{\sigma}
\newcommand{\thrs}{\tau}
\newcommand{\eat}{\eta}
\newcommand{\dptolerance}{\epsilon}
\newcommand{\meddev}{m}
\newcommand{\granul}{\gamma}
\newcommand{\numpatches}{n}
\newcommand{\numbuses}{\beta}

In this section, we introduce the algorithm to generate a fully parameterised stochastic model from the data. The algorithm consists of three main steps: map generation, patch identification, and model checking. 
Each step consists of a number of subroutines as displayed in Algorithm~\ref{alg: main algorithm}. The reasoning behind the subroutines is to make the code modular, so that individual techniques can be replaced should the need arise. In the following we present a brief overview of the program code --- note that an implementation written in Java is available online (see Section~\ref{sec: introduction}).

\begin{algorithm} 
\caption{}  
\label{alg: main algorithm}
\begin{algorithmic}[1]
	\Statex \vspace{-0.15cm} \;\; \textit{Map generation:}
    \State \texttt{ File heatMapFile $\leftarrow$ createObservationHeatMap(dataFile, $\intp$, $b$);} \label{it: heat map}
		\State \texttt{ File blurredMapFile $\leftarrow$ gaussianBlur(heatMapFile, $\gauss$);}
		\State \texttt{ File skeletonMapFile $\leftarrow$ skeletonise(blurredMapFile, $\thrs$, $\eat$);}
		\State \texttt{ File crossingsMapFile $\leftarrow$ detectCrossings(skeletonMapFile);}
		\State \texttt{ File graphFile $\leftarrow$ createPrunedGraphMap(crossingsMapFile, $\dptolerance$, $\meddev$);}
		\State \texttt{ File endpGraphFile $\leftarrow$ identifyEndPoints(graphFile);}
		\Statex \;\; \textit{Patch Identification:}
		\State \texttt{ int[\,] a $\leftarrow$ obtainObservationCounts(dataFile, endpGraphFile, $\granul$);} \label{it: obs count base}
		\State \texttt{ int[\,] j $\leftarrow$ jenksCluster(a, $\numpatches$);} \label{it: jenks}
		\Statex \;\; \textit{Model Checking \& Simulation}
		\State \texttt{ List$<$List$<$Integer$>$$>$ o $\leftarrow$ obtainObservations(dataFile, j);} \label{it: obs count jenks}
		\State \texttt{ double[][] erlangPars $\leftarrow$ fitErlangDistribution(o);} \label{it: erlang fit}
		\State \texttt{ BusSimulator simulator $\leftarrow$ new BusSimulator(erlangPars, $\numbuses$);} \label{it: simulator}
\end{algorithmic}
\end{algorithm}

Most of the steps in Algorithm~\ref{alg: main algorithm} require manually-specified parameters as input, as we  discuss in more detail in the next two sections. 
These parameter choices primarily concern the boundary between valid measurements and noise, which are best judged by a human user.
For example, the degree of noise and the existence of GPS ``shadows'' influence the choice of the parameters used in the first four steps. 
Although some level of input from a human planner is still required, the results of the subroutines provide a convenient way to provide feedback after each step has been completed. Also, for a single dataset (e.g., Seattle) a single value for each parameter will typically have good performance throughout the dataset, so parameter selection is not something that necessarily has to be repeated for every service.

In sections~\ref{sec: map generation}~to~\ref{sec: model checking}, we discuss the main steps of the algorithm in detail. However, before we continue, we first address the question of how to determine, for  an AVL measurement, the corresponding patch, as this question arises in multiple steps (\emph{e.g.}, in the \texttt{obtainObservationCounts} and \texttt{obtainObservations} routines in lines~\ref{it: obs count base}~and~\ref{it: obs count jenks} of Algorithm~\ref{alg: main algorithm}, respectively).
Since the bus GPS measurements are given as coordinates in a 2-dimensional continuous space, the most straightforward characterisation of patches is as polygons on the city map. This was indeed the approach taken in other papers such as \cite{reijsbergen2015patch} and \cite{reijsbergen2015probabilistic}. In this setting, an observation of a patch crossing time is collected from the data by recording the time of each measurement where a bus leaves a patch, and subtracting the time when it entered. To increase the accuracy of the crossing time, a linear interpolation between the measurements can be used. In this paper, we take a different approach: we assume that during the period of the day that we are interested in, buses do not deviate from their route. (This is not always valid, but we discuss approaches to mitigate resulting errors later on.) Consequently, given a bus location measurement, it is always possible to determine what fraction of the route has been completed and what fraction has yet to be done. This allows us to assign to each measurement a \emph{route completion percentage}, represented by a value in $[0, 1)$, where a value of $0$ represents the bus being at the beginning of the route, and values close to $1$ that the full route is about to be completed. The patches are then intervals that are subsets of $[0,1)$.

Calculating route completion percentages is \emph{not} straightforward in general, but amenable to automation as we will argue below. When a route is largely linear (\emph{e.g.}, the Airlink route in Edinburgh, see Fig.~\ref{fig: full heat map}), one possibility is to use the distance between the current location and the most recently visited terminus (\emph{i.e.}, start or end point). However, this is clearly inadequate when the route shape is more complex; see, for example, the Bellevue Express in Seattle (see Fig.~\ref{fig: route_map_550}).
To avoid users having to manually identify shape-defining corners in the route, we use the automatic procedure proposed by Biagioni \emph{et al.} \cite{biagioni2012map} to parse the data into a representation of the route in terms of a graph, \emph{i.e.}, a collection of nodes and edges. We then determine which of these edges are most likely to contain route termini, and then determine the main route segments using the most frequently occurring sequences of edges going from one terminus to another. If we know the edge sequences forming the routes, we project each measurement onto the nearest edge and calculate based on this information and the last terminus visited what the current route completion is. 

\subsection{Map Generation} \label{sec: map generation}

The first step of the algorithm is informed by the procedure from Biagioni \emph{et al.} \cite{biagioni2012map}, which we briefly summarise in this section. The first subroutine creates a heat map of GPS observations as displayed in Fig.~\ref{fig: full heat map}. A potential obstacle, which we for example observed near the airport and Waverley Station termini in Edinburgh, is that there can be areas on the route in which buses are never observed. Two possible explanations for this are GPS ``shadows'' --- where tall buildings or other structures block the GPS signal --- and overly aggressive data pre-processing. The gap near the Waverley Station end point is displayed in Fig.~\ref{fig: thumbnail1}. We compensate for the gap by interpolating between subsequent measurements --- see Fig.~\ref{fig: thumbnail2}. This procedure can be tuned using the parameter $\intp$, which determines the weight of interpolations compared to actual observations (setting $\intp = 0$ turns off the interpolation). Additionally, the contrast boost parameter $b$ informs a post-processing step that accentuates certain noise --- the only difference with $\delta$ is that its effect is non-linear.

\newcommand{\tnscale}{0.3\textwidth}
\newcommand{\circsize}{0.13cm}

\begin{figure}[!ht]
	\begin{center}
				
				\def\xx{-2.85}
				\def\xxx{6.5}
				\def\yy{-0.62}
				\def\dx{0.063}
				\def\dy{-\dx}
				\adjustbox{max width=\linewidth}{
		\begin{tikzpicture}[->,>=stealth',shorten >=1pt,auto,node distance=3cm,semithick, scale=0.8]					
			\node at (0,0) (s0) {
				\begin{tabular}{ccc}
					 \subfloat[]{\framebox{\includegraphics[width=\tnscale]{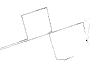}} \label{fig: thumbnail1}}
				 & \subfloat[]{\framebox{\includegraphics[width=\tnscale]{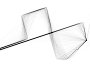}} \label{fig: thumbnail2}}
				 & \subfloat[]{\framebox{\includegraphics[width=\tnscale]{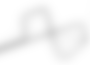}} \label{fig: thumbnail3}} \\
						 \subfloat[]{\framebox{\includegraphics[width=\tnscale]{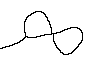}} \label{fig: thumbnail4}}
				 & \subfloat[]{\framebox{\includegraphics[width=\tnscale]{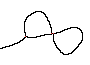}} \label{fig: thumbnail5}}
				 & \subfloat[]{\framebox{\includegraphics[width=\tnscale]{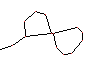}} \label{fig: thumbnail6}}
				\end{tabular}
			};
		\end{tikzpicture}
		}
	\end{center}
	\caption{Map thumbnail images for the portion of the Airlink route that intersects Princes Street. Specifically, these are: (a) uninterpolated, (b) interpolated, (c) blurred, (d) skeletonised, (e) skeletonised with the `crossings' identified (note that three pixels have been made red), and (f) the final graph representation with red pixels representing nodes and black pixels representing edges.}
	\label{fig: thumbnails}
\end{figure}

The next subroutine applies a Gaussian blur, as displayed in Fig.~\ref{fig: thumbnail3} - the intensity of the blur is determined by its standard deviation $\sigma$. This is followed by a filter in which pixels that are lighter than a certain threshold $\thrs$ are made completely white. We then find a skeleton representation of the remaining non-white pixels, as displayed in Fig.~\ref{fig: thumbnail4}. The procedure that has been implemented in the code is slightly different from the procedure described by Biagioni \emph{et al.} - instead of setting pixels at the end to white at each step of the skeletonisation procedure, we ``eat'' away at the edges by reducing the darkness of edge pixels by a certain degree $\eat$. This means that infrequently-occurring interpolations that ``cut corners'' are removed quicker. Low values of $\eat$ means that the algorithm takes longer to complete, but that it may be better able to get rid of certain types of noise (namely the ``hairs'' of the type also reported in \cite{davies2006scalable}).

Having obtained the skeleton map, we turn this into a graph as follows. First, we determine the locations of all the ends and ``crossings'' (specifically, all non-white pixels that do not have 0 or 2 non-white pixels in their 8-neighbour Von Neumann neighbourhood \cite{shi2009automatic}). These are the red pixels in Fig.~\ref{fig: thumbnail5}. We then determine the sequences of non-white pixels connecting these ends/crossings. For each such sequence, an initial graph representation is one where each pixel has a vertex and there is an edge for each pair of pixels that are Von Neumann neighbours. This graph is then pruned using the Ramer-Douglas-Peucker algorithm \cite{douglas1973algorithms} with tolerance $\dptolerance$. This transforms a sequence of pixels into a much smaller set of edges of different lengths, such that the rough 2-dimensional shape of the sequence is preserved. In order to remove extreme disparities between the edge lengths, we partition all edges that are longer than twice the median edge length into chunks roughly the size of the median divided by $\meddev$ (this aids the process of finding termini as described below). The final graph is then constructed by merging the vertices corresponding to the pixels which represent crossings, completing the algorithm described in \cite{biagioni2011easytracker}. The result for the Airlink route in Edinburgh's city centre is displayed in Fig.~\ref{fig: thumbnail6}.

In the resulting graph, the route can be represented by a sequence of directed edges. To determine the route, we first identify the edges containing the termini --- we assume that those are the edges in which a large amount of time is spent. We calculate for each (directed) edge how much time is spent on it on average, divided by the length of the edge. If this value is very high, then this indicates the presence of a terminus. In particular, we assume that the service has two termini (this can be generalised if the need exists) and hence select the two edges with the highest values. An alternative approach is to select the two extreme edges if the graph consists of a sequence of edges. Next, we determine the edge sequences that occur most often for buses travelling between the termini --- these are the route segments. For example, the Airlink has two segments, namely the West-East direction and the East-West direction. The length of the \emph{total} route starting from a specific start point (it does not matter which one is chosen) is then the sum of the edges leading back to the start point, if we take into account the direction of travel on the edges.

Given the graph map and route (segment) information, the route completion can be computed as follows: given a GPS measurement, we find the edge that it is closest to and which direction this bus is going (for this we use the previously visited terminus). This then maps uniquely to a part of the route --- buses are assumed to cross each directed edge at most once during the route. The distance to the start point can be found through the edge distances, and when this is divided by the total route length, this results in the required route completion measurement. Note that the length of the route can be different in the two directions: going by the edge lengths in the graph, the lengths of the west-east and east-west directions are roughly 12.4 kilometres and 12.3 kilometres respectively, whereas for the Bellevue Express in Seattle the route is roughly 20.2 kilometres long in both directions.

\subsection{Patch Identification} \label{sec: patches}

In the first subroutine of the second step, we compute for each bus GPS measurement the degree to which the bus has completed the route at the time of measurement, as a number in [0, 1).
 We then construct an initial patch structure by dividing the interval [0,1) into $\granul$ evenly-sized portions.
 We want to group together patches that are next to each other and that are ``similar'' in terms of the number of times buses were observed in the patches, since this is a good indication of the average bus speed in those patches. There are several ways to do this -- we have investigated two in more detail. The first method is to look for the pair of adjacent patches such that the merger of the two has the lowest number of bus observations. Merge these, and continue carrying out this step until the resulting patch has an observation count larger than the highest number of observations amongst the initial patches. As an advantage, this approach does not require that the final number of patches is specified \emph{a priori}.

For the second approach, we create a sequence of absolute differences between the number of observations in each patch and the next. We then use~\mbox{$K$-means} clustering in one dimension (this is called Jenks natural breaks optimisation) --- note that the $K$ in `$K$-means clustering' is part of its commonly used name, we instead denote the final number of patches by $\numpatches$. This is visualised in Fig~\ref{fig: obs count plot}, where the height of each bar represents the number of observations and the colours denote the final patch structure. In the remainder of this paper, we use this second approach.

The final patch structure for the Airlink service is displayed in Figures~\ref{fig: patches 0} and~\ref{fig: patches 1} for the west-east and east-west directions respectively. Here, the initial granularity $\granul$ was set to 50 and the final number of patches $\numpatches$ to 10. The two end points are part of the red patch on the left and the orange patch on the right respectively. Note that there is an asymmetry between the two directions: there are five patches between the two end patches on the west-east direction and only three on the east-west direction. 

\subsubsection*{Alternative Approaches}
Several other patch identification approaches have been considered, but were ultimately deemed less suitable than the aforementioned approach. We briefly discuss two of them in the following.

The first alternative approach is to do away with route completion percentages altogether, and take the full dataset of GPS measurements expressed only in terms of latitude/longitude coordinates. We then use $K$-means clustering to identify $K$ centre points of the clusters. A patch structure is then obtained using the Voronoi tessellation resulting from these $K$ centre points. An advantage of this approach is that it is easy to apply --- many tools support $K$-means clustering (\emph{e.g.}, the statistical package R). However, this approach is relatively crude and cannot handle ``complex'' route shapes, \emph{i.e.}, those that do not resemble a straight line.

The second is use the bus stops to demarcate the patches. To do this, we require a list of bus stops and their latitude/longitude coordinates. These locations can be inferred via the method of Biagioni \emph{et al.}\ \cite{biagioni2011easytracker}, or obtained via some other data source, \emph{e.g.}, the Lothian Buses data API for the Airlink route. 
Given the bus stops, we can then construct patches either using Voronoi tessellation or route completion percentages. However, the automated inference method of \cite{biagioni2011easytracker} is not always able to distinguish bus stops from busy junctions, and a succession of busy junctions in a densely populated area (\emph{e.g.}, in area around the eastern terminus of the Bellevue Express in Seattle) could lead to the creation of a multitude of spurious small patches. Whilst these could be merged, the question would then be what advantage there still is over the approach described above despite the additional effort.

\subsection{Model Checking and Simulation} \label{sec: model checking}

Once the patch structure has been obtained, the next step is to identify the probability distribution of the time spent in each of the $n$ patches. We focus on the Erlang and the hyper-Erlang distributions, which we validate through a comparison to other distributions in Section~\ref{sec: results}. We assume that we are only interested in measurements during given periods of interest during which the behaviour of vehicles is assumed to be similar. In our experiments we have found that, as expected, patch crossing times are noticeably different during rush hours, midday, and the night. The same is true for, \emph{e.g.}, weekends and weekdays. This is not only true for patch crossing times, but also for the schedule and model parameters that are input by the user such as $\beta$, the number of buses assigned to the service. Note that this does not affect the generality of the method: for example, if one wants to know whether a bus service has adequate performance throughout the day, one should partition the dataset and check whether the performance requirements are satisfied in each of the portions. In fact, the parameter fitting procedures of this paper are helpful to inform such a partitioning: if the parameters are comparable between time segments then they can justifiably be merged. 

The parameters of the Erlang and hyper-Erlang distributions are found by matching them to the patch crossing times observed in the data. Patch crossing time observations are obtained in the following way: we calculate route completion percentages for each measurement in the dataset that occurs during a period of interest, for all buses that service the selected route. To increase the granularity of the data, we perform a linear interpolation between the measurements for each bus. That is, for every measurement we also calculate its UNIX time, \emph{i.e.}, the number of seconds since 1 January 1970. For each measurement $i$, let $t_i$ be its Unix time, $r_i$ its route completion percentage, and $t_{i+1}$ and $r_{i+1}$ the Unix time and route completion percentage of the next measurement of the \emph{same bus}. We then create interpolated measurements as follows: let $\Delta t = t_{i+1} - t_{i}$ and $\Delta r = r_{i+1} - r_{i}$, then we create new measurements with times $ t_i+1, t_i + 2 \ldots, t_i + \Delta t - 1$ and route completion percentages $r_i + \frac{1}{\Delta t}\Delta r, r_i + \frac{2}{\Delta t}\Delta r, \ldots, r_{i} + \frac{\Delta t - 1}{\Delta t} \Delta r$. We interpolate in terms of route completion percentage instead of latitude/longitude to increase measurement accuracy --- \emph{e.g.}, interpolation in space can lead to corners being cut short, whereas interpolation in route completion compels the buses to adhere strictly to the route. 

For each bus, we keep track of the number of seconds since it crossed into its current patch. If the bus crosses the boundary between patches~$j$~and~$j+1$, or between between patches~$n-1$~and~0, then we record the measurement time minus the time at which the bus crossed into patch $j$ as a crossing time observation for patch~$j$. Due to measurement errors, it may happen that a bus appears to move backwards, i.e., cross from patch $j$ into $j-1$ or from $0$ into $n-1$.\footnote{For example, this appears to happens for bus 5019 in the Seattle dataset, between 16:14:30 and 16:21:53 on 1 November 2001.} This may lead to spurious small crossing time observations, which may have a large distorting effect on the parameter fitting procedure. As such, if a bus crosses into the previous patch, the number of seconds since crossing into it is set to~$-1$ instead of $0$ to flag that this measurement is probably faulty, and will hence be ignored. This flag is also set if the time between measurements is more than 5 minutes --- to ignore the first measurement in a period of interest, and because interpolation becomes too crude if there is too much time between measurements --- or if a distance of more than 5 kilometres is travelled between measurements --- this is due to a quirk in the Seattle dataset where buses in the downtown area suddenly seem to appear in Bellevue, before reappearing downtown a few minutes later.\footnote{For example, this appears to happens for bus 5203 in the Seattle dataset, between 17:45:55 and 18:07:26 on 31 October 2001.}

Once a full dataset of observations $\vec{x} = (x_{j1}, x_{j2}, \ldots, x_{j\numobs(j)})$ has been created, where $\numobs(j)$ is the number of observations for patch $j$, we need to obtain the optimal Erlang and hyper-Erlang distribution parameters $k_j$ and $\lambda_j$. In our implementation, the procedure for the Erlang distribution is as follows: starting with $k_j = 1$, we set $\lambda_j = k_j / \bar{x}_j$ with $\bar{x}_j = \sum_{i=1}^{\numobs(j)} x_{ji} / \numobs(j)$, meaning that we set the mean of the Erlang distribution to equal the sample average (in statistics, this is called the `method of moments'). We then calculate the corresponding log-likelihood value 
\[
	l(\vec{x}; k_j, \lambda_j) = \sum_{i=1}^{\numobs(j)}  \log(f(x_{ji};k_j, \lambda_j))
\]
with $f$ equal to the function $f_S$ of \eqref{eq: erlang pdf}. We then increase $n_j$ by one, and repeat this procedure - we continue until we reach the value $k_j$ such that its log-likelihood is lower than in the previous step. We then choose the $\lambda_j$ and $k_j$ of the previous step as the distribution parameters. To obtain the hyper-Erlang results in Section~\ref{sec: results}, we have used the tool HyperStar \cite{reinecke2012hyperstar}, although it should be noted that this tool uses a randomised algorithm and that the found parameters typically vary between tries.

The Erlang/hyper-Erlang distributions describe the movement of the buses through the patches when they are servicing the route. During simulation, we initialise the buses as either starting at a single terminus, or uniformly along the route (for long-run simulations, the difference is negligible). We keep track of a single global clock, and for each bus its current patch and the number of completed phases within that patch. We then draw phase completion times for each bus using the rate in their current patches (and branches in case of the hyper-Erlang distribution). We then identify the bus with the next phase completion, increment its phase, potentially change its patch (and possibly select a hyper-Erlang branch) and draw its next phase completion time. We continue until the statistical model checking front-end (\emph{i.e.}, MultiVeStA) has found that enough samples have been drawn for the simulation to be completed.

In practice, the buses are regularly observed to wait at the termini to increase the regularity of the service. To enforce an implicit schedule on the buses, we enforce that buses cannot leave end patches unless $t > r d_i + h_{ij}$. Here, $t$ is the global time, $d_i$ is an integer that is incremented by $1$ every time the route is completed (\emph{i.e.}, the first patch is entered), $r$ is the \emph{timetabled} amount of time that is assigned to buses to complete the entire route, and $h_{ij}$ is a constant that ensures that different buses leave patch $j$ at different time points. A typical choice for $r$ is the total duration of the route (including time spent at the end points). Let the cumulative means $c_j$ be given as $c_j = \sum_{{\iota}=1 }^{j-1} \mu_{\iota}$ where the $\mu_{\iota}$ are the mean patch completion times. Then a typical choice for $h_{ij}$ equals $r (i-1) / \numbuses + c_j$, with $\numbuses$ the total number of buses, so that the times between bus departures are timetabled to all be the same. 

To illustrate these choices of $r$ and $h_{ij}$ for the Airlink case study, first note that the total route duration (including termini) for midday Airlink buses equals $5\,259$ seconds or $87.65$ minutes (this can be seen by summing the values of $\mu$ in Table~\ref{tab: patch parameters} in Section~\ref{sec: results}). In the dataset, 11 buses typically seem to be servicing the route in the midday. This means that $r = 5\,259$ and \mbox{$h_{ij} = 5\,259/11 (i-1) + c_j = 478.09 (i-1) + c_j$}, with $c_0 = 0$ and $c_6 = 2\,744$. This means that according to the timetable buses depart from the airport roughly every $478.09$ seconds. This is very close to the value of $477.2047$ that was reported in Table~3 of \cite{reijsbergen2014formal}. 

\subsubsection*{Performance Metrics}

To measure system performance, we use the metrics used by the Scottish government (see also \cite{reijsbergen2014formal}). These measures are only relevant for frequent services, defined as routes for which six or more bus arrivals are scheduled per hour. For frequent services, exact timetable performance is not as relevant as the regularity of the \emph{headway}, the time between subsequent bus departures (see, \emph{e.g.},\cite{moreira2015improving}). In particular, we consider the following metrics:

\begin{enumerate}
	\item the \emph{excess waiting time} (EWT),
	\item the \emph{extreme-value waiting time} performance (EVWT), and
	\item the \emph{buses-per-hour} performance (BPH).
\end{enumerate} 
The EWT is the average experienced waiting time minus the ``timetabled'' waiting time, where by the ``timetabled'' waiting time we mean the waiting time experienced by passengers arriving uniformly to the bus stop when there is \emph{zero} headway variance. In particular, with $\sigma^2$ denoting the headway variance and $\mu$ the headway mean, the excess waiting time can be shown to equal $\frac{1}{2}\sigma^2/\mu$. The EVWT is the probability that the amount of time between subsequent bus departures is more than 15 minutes. The BPH performance is the steady-state probability that fewer than six buses have departed in the previous hour. The Scottish government's requirement on the EVWT and BPH is that they should be at most 5\% at the starting point of a journey. 
The EWT should not exceed 75 seconds at a set of important bus stops called ``timing points''.

To evaluate the aforementioned metrics using statistical model checking, we introduce several additional variables during simulation. For each bus $i \in \{1,\ldots,\numbuses\}$ and patch $j \in \{1,\ldots,\numpatches\}$, we maintain a clock $z_{ij}$ that represents the time since bus $i$ entered patch $j$. For the BPH, we use the counter $H_j$ which for each patch $j$ denotes the number of buses that departed in the past hour. Hence, $ H_j = \sum_{i=1}^\numbuses {\bf 1}(z_{ij} < 3600)$, where ${\bf 1}(A)$ equals $1$ if the boolean expression $A$ is true and $0$ otherwise.  The clock $y_j$, which is used by both the EWT and EVWT, represents the time since the last visit by a bus to patch $j$, and equals $y_j = \min_{i\in\{1,\ldots,\numbuses\}} z_{ij}$. Finally, the counter $\ccounter_j$ is incremented every time a bus leaves patch $j$. 

\begin{table}
	\makebox[\textwidth][c]{
		\begin{tabular}{|l|l|}
		\hline
			\multicolumn{1}{|c|}{QuaTEx expression} & \multicolumn{1}{c|}{Property} \\ \hline
			\texttt{ewt() \ = 0.5 * (s.rval("y\_j") - mu\_tot)} & \texttt{ S [ ewt(), "}\texttt{\ccounter\_j" ] < 75} \\ 
			 \texttt{\ \ \ \ \ \ \ \ \ \ * (s.rval("y\_j") - mu\_tot) / mu\_tot;} & \\ \hline
			\texttt{evwt() = if \{s.rval("y\_j") > 900\}} & \texttt{ S [ evwt(), "}\texttt{\ccounter\_j" ] < 0.05 } \\ 
			\texttt{\ \ \ \ \ \ \ \ \ \ then 1 else 0 fi;} & \\ \hline
			\texttt{bph() \ = if \{s.rval("H\_j") < 6\}} & \texttt{ S [ bph(), "time" ] < 0.05} \\
			\texttt{\ \ \ \ \ \ \ \ \ \ then 1 else 0 fi;} & \\ \hline
		\end{tabular}
	}
	\caption{The QuaTEx expressions for the EWT, EVWT, and BPH.}
	\label{tab: quatex punctuality measures}
\end{table}

Given these variables, the three system performance metrics can be expressed using the language MultiQuaTEx (see Section~\ref{sec: model checking}) as given in Table~\ref{tab: quatex punctuality measures}. For the EWT, we use the value $\mu_{\text{tot}}$ which denotes the timetabled headway. Note that for the BPH the notion of time is given by the global clock, denoted by \texttt{"time"}, whereas for the EWT and EVWT we use $\ccounter_j$ to denote time. 
In Section~\ref{sec: results}, we will give numerical results involving these properties.


\section{Results} \label{sec: results}

In this section, we illustrate the applicability of the proposed method by means of two case studies. We begin with the Airlink service in Edinburgh in Section~\ref{sec: airlink current}. We confirm the experiments done in \cite{reijsbergen2014formal} that show that the Airlink has excellent performance.  In Section~\ref{sec: seattle}, we consider one of the services in Seattle, namely the Bellevue express. Its route covers downtown Seattle and its performance in 2001 was noticeably below that of the Airlink service. In Section~\ref{sec:strategies}, we have another look at the Airlink service --- our simulations, using a set-up with altered parameters, suggest that the average waiting time of passengers can still be improved by 1.5 minutes by using a strategy that involves buses adapting to each other's behaviour.

\subsection{The Airlink Service in Edinburgh} \label{sec: airlink current}

\subsubsection*{Patch Structure}

\newcommand{\alscale}{\textwidth}
\begin{figure}[!hp]
	\begin{center}
		\begin{tabular}{c}
			\subfloat[]{\makebox[\alscale][c]{\framebox{\includegraphics[width=\alscale]{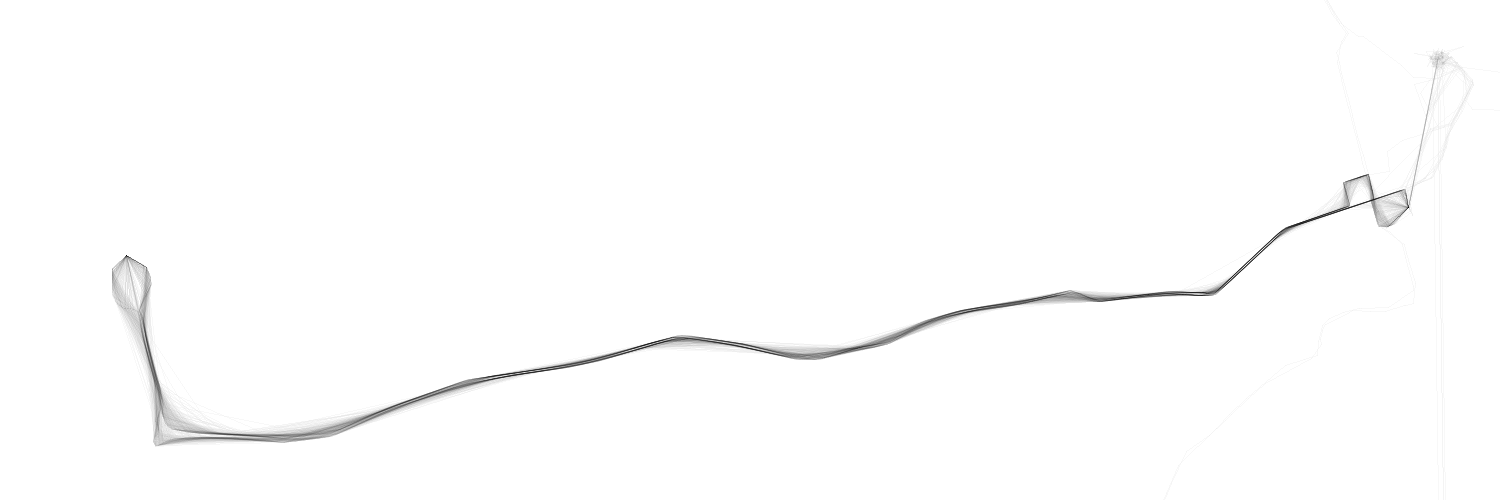}}} \label{fig: full heat map}} \\
			\subfloat[]{\makebox[\alscale][c]{\framebox{\includegraphics[width=\alscale]{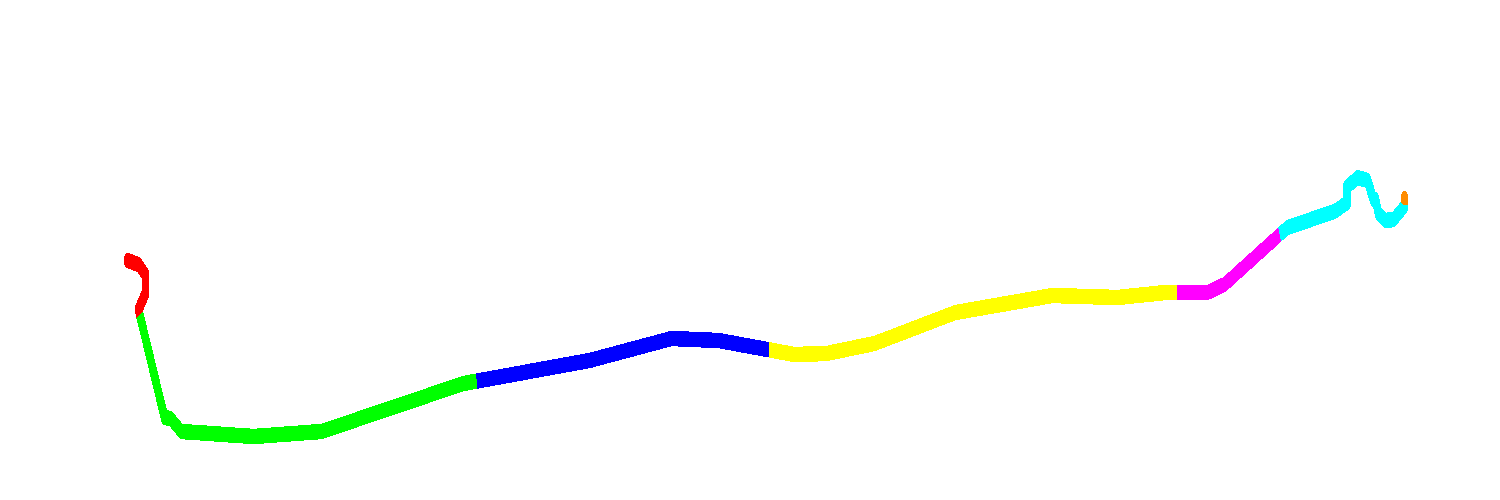}}} \label{fig: patches 0}} \\
			\subfloat[]{\makebox[\alscale][c]{\framebox{\includegraphics[width=\alscale]{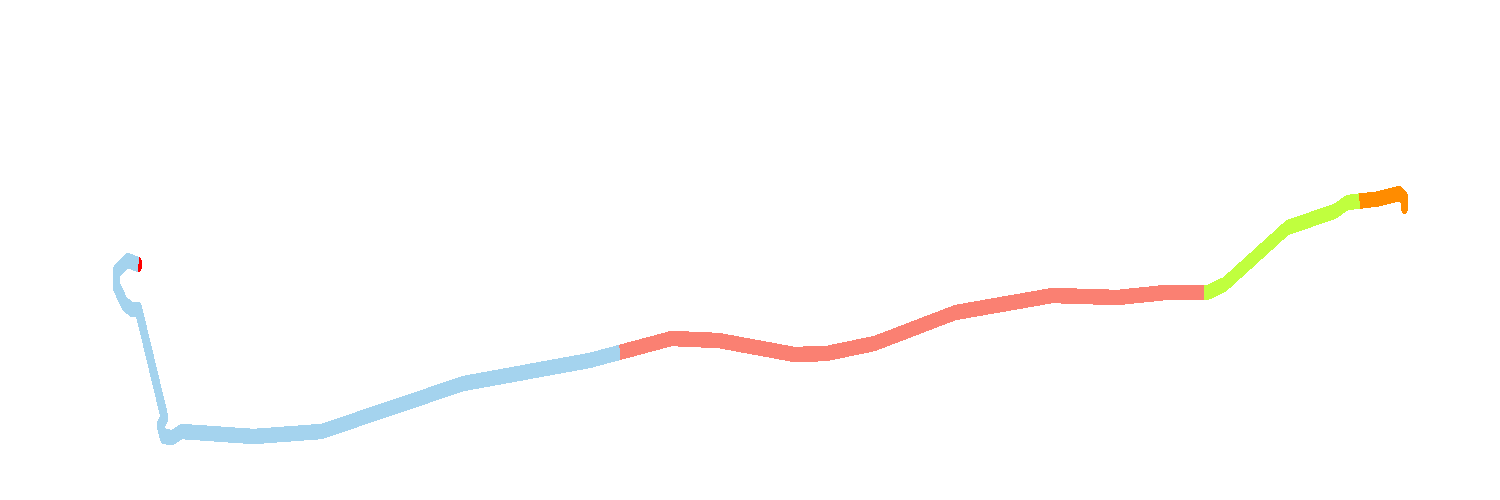}}} \label{fig: patches 1}}
		\end{tabular}
	\end{center}
\caption{Top: heat map of the full Airlink route, \emph{including} entries outside the 10AM-3PM window. Centre and bottom: automatically obtained patch structures for the west-east and east-west direction respectively.}
\end{figure}

The full heat map of the Airlink service, including interpolations, that is created by running the \texttt{createObservationHeatMap} function in line~\ref{it: heat map} of Algorithm~\ref{alg: main algorithm}, is displayed in Figure~\ref{fig: full heat map}. For Figure~\ref{fig: full heat map}, we have decided to display the heat map for the \textit{full dataset}, whereas we only consider the time period between 10AM and 3PM for all other Airlink figures and tables. We have done this because the full heat map shows the buses still reporting measurements whilst stationed at Lothian Buses' depot on Annandale Street, at the ``cloud'' in the top right corner of Figure~\ref{fig: full heat map}. It also visualises data quirks: the vertical lines going downward from the Annandale depot correspond to interpolations between bus depot measurements and unlikely measurements in other parts of UK, sometimes as far as Wales. The regularly appearing (and hence quite dark-coloured) straight line between the Annandale depot and the Waverley station terminus on the Easternmost part of the route is presumably due to the measurement device being off either after completion or before the start of the route.\footnote{As discussed in Section~\ref{sec: model checking}, such measurements are not considered when we determine patch crossing time observations.} 

\begin{figure}[!hpb]
	\begin{center}
			\vspace{-0.7cm}\makebox[\textwidth][c]{\includegraphics[width=0.9\textwidth]{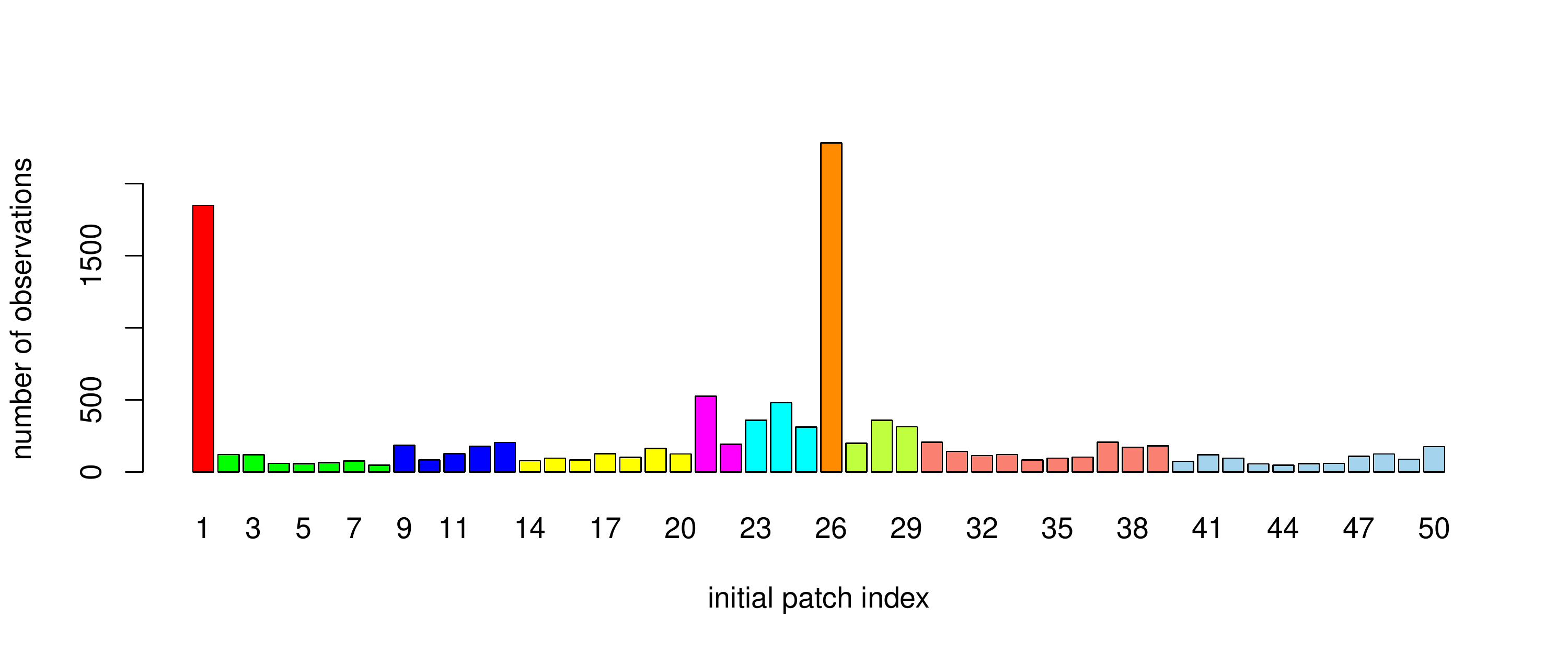}}\vspace{-0.7cm}
	\end{center}
\caption{In this bar chart, the height of each bar represents the number of bus observations (without interpolation) in the corresponding initial patch. The colouring indicates the patch structure resulting from Jenks natural breaks optimisation on the differences between subsequent bar heights. The colouring used here is identical to that used in Figures~\ref{fig: patches 0}~and~\ref{fig: patches 1}.}
 \label{fig: obs count plot}
\end{figure}

After the completion of steps 1-6 of Algorithm~\ref{alg: main algorithm}, we partition the route into 50 initial patches, and determine for each patch how many times a bus was observed in this patch. The resulting bar chart is displayed in Figure~\ref{fig: obs count plot}. After K-means clustering with 10 means, we obtain the patch structure as indicated by the colouring of the bars in Figure~\ref{fig: obs count plot}. A visual representation of the patches in given in Figures~\ref{fig: patches 0}~and~\ref{fig: patches 1}. Because the two route segments are not of equal length, and because entering the city centre is more time-consuming than leaving it, two patches are created for the city centre area, and only one in the opposite direction (excluding the terminus patch). In general, it can be observed that, as expected, buses spend more time in the densely populated areas in the eastern part of the route  than on Glasgow Road towards the west. 

\subsubsection*{Parameter Fitting}

In Table~\ref{tab: patch parameters}, we have displayed the Erlang parameter values $\lambda_j$ and $k_j$ for each patch $j=1,\ldots,10$. We also display for each patch its mean patch crossing time $\mu_j$ (in seconds), the standard deviation $\sigma_j$, and the coefficient of variation $c_{v,j}$. The coefficient of variation is defined as the ratio of the standard deviation to the mean --- it is noticeably higher for inner-city patches than for the patches in the city outskirts. In particular, the patch near Haymarket station (patch 5) has a very high coefficient of variation. 

\definecolor{red}{RGB}{255, 0, 0}
\definecolor{green}{RGB}{0, 255, 0}
\definecolor{blue}{RGB}{0, 0, 255}
\definecolor{yellow}{RGB}{255, 255, 0}
\definecolor{magenta}{RGB}{255, 0, 255}
\definecolor{cyan}{RGB}{0, 255, 255}
\definecolor{darkorange}{RGB}{255, 140, 0}
\definecolor{olivedrab1}{RGB}{192, 255, 62}
\definecolor{salmon}{RGB}{250, 128, 114}
\definecolor{lightskyblue2}{RGB}{164, 211, 238}
\definecolor{khaki3}{RGB}{205, 198, 115}
\definecolor{lavenderblush3}{RGB}{205, 193, 197}
\begin{table}
	\makebox[\textwidth][c]{
		\begin{tabular}{c|cccccccccc}
			Patch & 1 \clbox{red} & 2 \clbox{green} & 3 \clbox{blue} & 4 \clbox{yellow} & 5 \clbox{magenta} & 6 \clbox{cyan} & 7 \clbox{darkorange} & 8 \clbox{olivedrab1} & 9 \clbox{salmon} & 10 \clbox{lightskyblue2}\\\hline
			$l$ & 0.49 & 3.46 & 2.47 & 3.46 & 0.99 & 1.48 & 0.49 & 1.48 & 4.94 & 5.43\\
			$k$ & 44 & 106 & 68 & 73 & 17 & 37 & 40 & 30 & 78 & 101\\
			$\lambda$ & 0.0482 & 0.4190 & 0.1858 & 0.2011 & 0.0523 & 0.0710 & 0.0419 & 0.0765 & 0.1196 & 0.1895\\
			$\mu$ & 912.0 & 253.0 & 366.0 & 363.0 & 325.0 & 521.0 & 954.0 & 392.0 & 652.0 & 533.0\\
			$v$ & 1.95 & 49.20 & 24.29 & 34.29 & 10.94 & 10.24 & 1.86 & 13.61 & 27.27 & 36.70\\
			$\sigma$ & 137.49 & 24.57 & 44.38 & 42.49 & 78.82 & 85.65 & 150.84 & 71.57 & 73.82 & 53.04\\
			$c_v$ & 0.1508 & 0.0971 & 0.1213 & 0.1170 & 0.2425 & 0.1644 & 0.1581 & 0.1826 & 0.1132 & 0.0995
		\end{tabular}
	}
	\caption{Patch parameters for the Airlink service, midday (10AM - 3PM). For each patch, we display the length $l$ of the patch in kilometres, Erlang shape and scale parameters $k$ and $\lambda$, the mean $\mu = k / \lambda$ in seconds, the average bus speed $v$ in the patch in kilometres per hour, the standard deviation $\sigma = \sqrt{k}/\lambda$, and the coefficient of variation $c_v = 1 / \sqrt{k}$.}
	\label{tab: patch parameters}
\end{table}

Empirical CDF plots for the patch crossing time observations are displayed in Figure~\ref{fig: cdf plots} in the Appendix. As we can see, the Erlang distributions have a good fit. To evaluate the goodness-of-fit numerically, we use the Anderson-Darling test to test the null hypothesis that a sample is drawn from the fitted distribution. Of course, the test can only be used to disprove the null hypothesis, whereas ideally we would like to prove it. Moreover, the test is biased against disproval in our setting, because the test assumes that the sample is drawn independently from the null hypothesis distribution --- this is not the case, as the parameters of the distribution where calculated using the same sample. However, if the Anderson-Darling test is still able to reject the null hypothesis, despite the bias against doing so, then we can take this as evidence that the fitted distribution is inappropriate. The test statistics and $p$-values have been calculated using the \texttt{goftest} package of R.

\setlength{\tabcolsep}{6pt} 
\begin{table}[!ht]
	\centering
	\adjustbox{max width=\linewidth}{
		\begin{tabular}{c|cccccccccc}
			Patch & 1 \clbox{red} & 2 \clbox{green} & 3 \clbox{blue} & 4 \clbox{yellow} & 5 \clbox{magenta} & 6 \clbox{cyan} & 7 \clbox{darkorange} & 8 \clbox{olivedrab1} & 9 \clbox{salmon} & 10 \clbox{lightskyblue2}\\\hline
N. obs. &  71 & 80 & 80 & 79 & 79 & 76 & 75 & 80 & 79 & 75 \\ 
 AD &  0.777 & 1.1076 & 0.5353 & 0.4338 & 0.6865 & 0.108 & 0.843 & 0.6958 & 0.8432 & 0.3954 \\ 
$p$-value &  0.4974 & 0.3053 & 0.7105 & 0.8143 & 0.5697 & 0.9999 & 0.4506 & 0.5618 & 0.4505 & 0.8529 \\ 
		\end{tabular}
	}
	\caption{For each of the 10 patches for the Airlink service, the number of patch crossing time observations, Anderson-Darling test statistics, and corresponding $p$-values. The $p$-values are relatively high in all cases.}
	\label{tab: airlink gof}
\end{table}
\setlength{\tabcolsep}{6pt} 

We have displayed the test statistics and $p$-values in Table~\ref{tab: airlink gof}. The $p$-values are very high, which also suggests that the Erlang distribution has a good fit for the Airlink's patch crossing time distributions. The choice for a very specific observation period (\emph{i.e.}, between 10AM and 3PM, on a Tuesday, Wednesday and Thursday) may contribute to the regularity of the measurements. In Section~\ref{sec: seattle}, we will see that the Erlang distribution can have a less good fit when conditions are more challenging (\emph{i.e.}, during rush hour).

\subsubsection*{Performance Evaluation}

To demonstrate the usefulness of our method, we can use it to reproduce the results from \cite{reijsbergen2014formal}. We use the three performance metrics for frequent services used in \cite{reijsbergen2014formal} and discussed in Section~\ref{sec: model checking}: namely the EWT, the EVWT, and the BPH.\footnote{Note that \cite{reijsbergen2014formal} considered two versions of the BPH: namely the Steady-State Buses-per-Hour Requirement (SSBHR) and the Day-Long Buses-per-Hour Requirement (DLBHR). The reason was that the requirement specified in the Scottish governments BPIPS document \cite{bpips} could be interpreted in different ways. The BPH here uses the underlying metric of the SSBHR as we feel that this is the more natural interpretation of the requirement.} In \cite{reijsbergen2014formal}, these metrics were computed using an elementary (ARMA) time series model that did not consider the movement of buses through space. 
Before we present a comparison of headway correction strategies, we will first determine that the results coincide.

\setlength{\tabcolsep}{0pt} 
\begin{table}
	\makebox[\textwidth][c]{
		\begin{tabular}{c|cccccccccc}
			Patch & 1 \clbox{red} & 2 \clbox{green} & 3 \clbox{blue} & 4 \clbox{yellow} & 5 \clbox{magenta} & 6 \clbox{cyan} & 7 \clbox{darkorange} & 8 \clbox{olivedrab1} & 9 \clbox{salmon} & 10 \clbox{lightskyblue2}\\\hline
			EWT\; & \perr{0.25}{0.02} & \perr{1.51}{0.04} & \perr{5.84}{0.13} & \perr{9.12}{0.24} & \perr{22.45}{0.56} & \perr{36.39}{0.78} & \perr{0.25}{0.01} & \perr{10.82}{0.28} & \perr{22.57}{0.64} & \perr{28.14}{0.75}\\
			EVWT\; & ---  & ---  & ---  & ---  & \perr{0.002}{0.00} & \perr{0.012}{0.00} & ---  & ---  & \perr{0.002}{0.00} & \perr{0.004}{0.00}\\
			BPH\; & ---  & ---  & ---  & ---  & ---  & \perr{0.000}{0.00} & ---  & ---  & ---  & --- 
		\end{tabular}
	}
	\caption{For each of the 10 patches for the Airlink service, its performance expressed in the form of the three metrics discussed in Section~\ref{sec: model checking}. The dashes mean that the corresponding event (\emph{e.g.}, more than 15 minutes between subsequent bus arrivals) was not observed during the simulation, resulting in an estimate of 0. The entries saying ``0.000'' mean that the event of interest was observed, but that the resulting probability was still rounded down to zero.}
	\label{tab: airlink current}
\end{table}
\setlength{\tabcolsep}{6pt} 

In Table~\ref{tab: airlink current}, we display estimates of the current EWT, EVWT, and BPH for each of the 10 patches in the Airlink model. We drew as many simulations as needed (using sequential hypothesis testing) to reach a conclusion on whether the requirement was met (\emph{i.e.}, 75 seconds for the EWT etc.), and additionally that the relative confidence interval half-width had to be at most 10\% --- this was to ensure that we would get reasonably accurate confidence intervals even when the requirements are comfortably met. We left the entry blank if we had not yet observed the event of interest after 300 seconds of simulation. A complication in many settings is that the Airlink service is so reliable that observing serious headway deviations is very unlikely, in particular at the beginning of the route. If we do not observe the event of interest, we cannot estimate its probability --- the so-called \emph{rare-event} problem. Since this means that the requirements are satisfied in our setting, this is not a complication. 

Note that in Table~\ref{tab: airlink current}, the service typically starts to perform worse when the buses get further from the termini. This agrees with the results presented in \cite{reijsbergen2014formal}. The EWT values are somewhat lower than were reported in \cite{reijsbergen2014formal}. For example the EWT for patch 1 in Table~\ref{tab: airlink current} is 0.25 seconds, whereas the value of 7.9773 seconds is reported in Table~3 of \cite{reijsbergen2014formal} as the EWT at the airport bus stop. This is to be expected: in the model a bus will leave an end stop the very second the timetable says that it should, whereas  human drivers will not be similarly precise. This effect wanes as the route progresses: at the end of Patch 3, the difference is 5.84 versus 17.5301 seconds. At the end of patch 6, the difference is 36.39 versus 35.5752 seconds. Do note that the results in the tables are not entirely comparable since different time intervals were used: 9AM-5PM in \cite{reijsbergen2014formal} and 10AM-3PM in this paper. The EVWT at the end of Patch 6 is roughly 1.2\%, which is inside the confidence interval reported in Table~4 of \cite{reijsbergen2014formal}. The BPH is close to zero across the route, which is also consistent with Table~6 of \cite{reijsbergen2014formal}.

It is clear from Table~\ref{tab: airlink current} that the Airlink service meets the requirements set by government regulators. The EWT remains below 75 seconds at each of the patches; similarly, the EVWT and BPH remain below 5\% even at parts in the middle of the route where the requirements no longer apply. However, there is still room for improvement, as we will see in Section~\ref{sec:strategies}. However, before we move on to bus performance improvement strategies for the Airlink, we present the modelling results for the Seattle dataset.

\subsection{The Bellevue Express in Seattle} \label{sec: seattle}

\subsubsection*{Patch Structure}

One of the main strengths of the approach introduced in this paper is its generality: under mild assumptions, any transport service for which AVL measurements have been collected over a prolonged period can be analysed in the same manner. To demonstrate this generality, we apply our method to the publicly dataset involving buses in Seattle \cite{rice2003data}, and in particular on the Bellevue Express, which is/was a frequent service during evening rush hours (\emph{i.e.}, between 4PM and 6PM on weekdays).  
\begin{figure}[ht!]
\begin{center}
	\begin{tabular}{cc}
		\framebox{\includegraphics[width=0.6\textwidth]{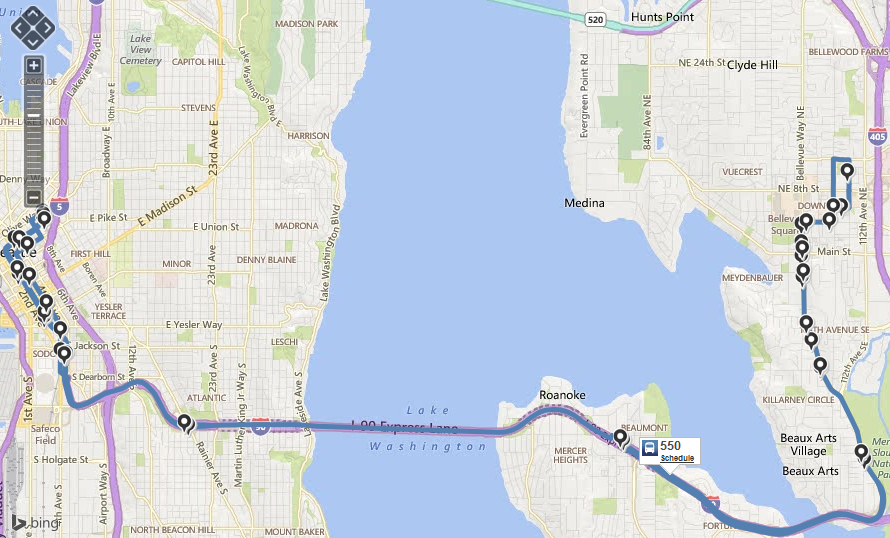}} 
	\end{tabular}
\end{center}
\caption{Screenshot from \url{http://www.soundtransit.org/schedules/ST-Express-Bus/550/map} {displaying the current (as per July 2016) route of SoundTransit Route 550}, the Bellevue Express.}
\label{fig: route_map_550}
\end{figure}
The route map of the Bellevue express as of July 2016 is displayed in Figure~\ref{fig: route_map_550}. As we can see in Figure~\ref{fig: skeleton550}, the route has only slightly been changed in the more than 15 years since 2001, when the data was recorded -- the only major difference is in the North-East part of the route. There are no gaps in the measurements similar to what we observed for the Airlink data, which means that it is not necessary to put much weight on interpolations between subsequent measurements. 

\begin{figure}[!th]
\begin{center}
\adjustbox{max width=\linewidth}{
	\begin{tabular}{cc}
		\multicolumn{2}{c}{\subfloat[]{\framebox{\includegraphics[width=0.9\textwidth]{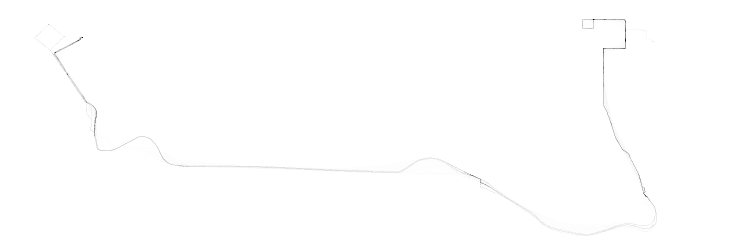}} \label{fig: skeleton550}}}
		\\
		\subfloat[]{\framebox{\includegraphics[width=0.45\textwidth]{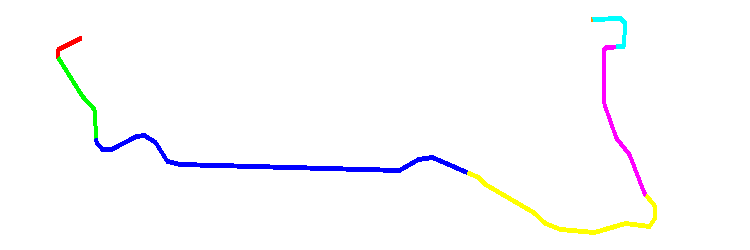}} \label{fig: patches550westeast}} &
		\subfloat[]{\framebox{\includegraphics[width=0.45\textwidth]{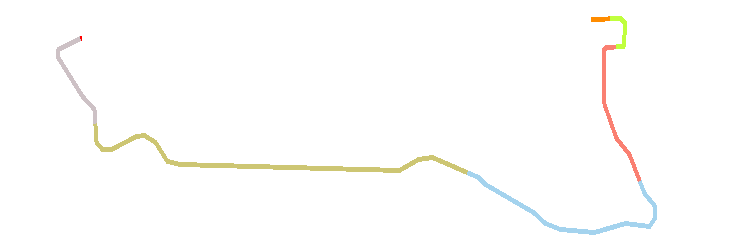}} \label{fig: patches550eastwest}}
	\end{tabular}
	}
\end{center}
\caption{Heat map, and patch structure in both directions for the Bellevue Express in Seattle.}
\label{fig: patches_550}
\end{figure}

As measured by the graph map produced in line 5 of Algorithm~\ref{alg: main algorithm}, the route is roughly 20.16km in both directions. We use 12 patches instead of the 10 for the Airlink as this led to better fitting results. During rush hour, the Bellevue Express buses spend little time waiting at the western terminus, sometimes passing it completely in the time between subsequent measurements. (This is also evident from the CDF plot for patch 1 in Figure~\ref{fig: cdf plots seattle} in the Appendix, wich includes very short and very long crossing times.) This complicates the procedure for determining the termini, however in this case the two edges on the far ends of the graph are obvious candidates.

\subsubsection*{Parameter Fitting}

The parameters characterising the patches are displayed in Table~\ref{tab: seattle patch parameters}. The fact that buses spend a considerably larger amount of time in the terminus in the east (Patch 7) than in the west (Patch 1) is evident from the corresponding values for $\mu$. The average speed in Patch 3, which corresponds to the highway crossing Mercer Island, is 60.48 km/h, which is much higher than any patch in the Airlink dataset. However, the patches for the west-east direction have higher average speeds than those for the east-west direction. Another interesting feature is Patch 8, corresponding to an area in Bellevue where an average speed (including time spent waiting at stops and junctions) of only about 6.68 km/h is observed. In general, the coefficients of variation in Table~\ref{tab: seattle patch parameters} are much higher than in Table~\ref{tab: patch parameters}, its counterpart for the Airlink service.

	\setlength{\tabcolsep}{3.5pt} 
\begin{table}
	\makebox[\textwidth][c]{
		\begin{tabular}{c|cccccccccccc}
			Patch & 1 \clbox{red} & 2 \clbox{green} & 3 \clbox{blue} & 4 \clbox{yellow} & 5 \clbox{magenta} & 6 \clbox{cyan} & 7 \clbox{darkorange} & 8 \clbox{olivedrab1} & 9 \clbox{salmon} & 10 \clbox{lightskyblue2} & 11 \clbox{khaki3} & 12 \clbox{lavenderblush3}\\\hline
			$l$ & 0.67 & 2.02 & 7.73 & 4.70 & 3.70 & 1.34 & 0.34 & 1.01 & 3.36 & 5.04 & 8.06 & 2.35\\
			$k$ & 2 & 46 & 155 & 45 & 38 & 24 & 1 & 24 & 20 & 14 & 28 & 20\\
			$\lambda$ & 0.0044 & 0.1098 & 0.3370 & 0.1275 & 0.0784 & 0.0538 & 0.0010 & 0.0442 & 0.0361 & 0.0284 & 0.0482 & 0.0362\\
			$\mu$ & 453.0 & 419.0 & 460.0 & 353.0 & 485.0 & 446.0 & 984.0 & 543.0 & 554.0 & 493.0 & 581.0 & 552.0\\
			$v$ & 5.34 & 17.32 & 60.48 & 47.97 & 27.43 & 10.85 & 1.23 & 6.68 & 21.83 & 36.80 & 49.97 & 15.34\\
			$\sigma$ & 320.32 & 61.78 & 36.95 & 52.62 & 78.68 & 91.04 & 984.00 & 110.84 & 123.88 & 131.76 & 109.80 & 123.43\\
			$c_v$ & 0.7071 & 0.1474 & 0.0803 & 0.1491 & 0.1622 & 0.2041 & 1.0000 & 0.2041 & 0.2236 & 0.2673 & 0.1890 & 0.2236
		\end{tabular}
	}
	\caption{Same as Table~\ref{tab: patch parameters}, but for the 12 patches of the Bellevue Express.}
	\label{tab: seattle patch parameters}
\end{table}
\setlength{\tabcolsep}{6pt} 

The CDFs of the patch crossing time observations, including fitted Erlang distributions, are displayed in Figure~\ref{fig: cdf plots seattle} in the Appendix. The Anderson-Darling test statistics and $p$-values for the Bellevue Express are displayed in Table~\ref{tab: seattle gof}. We can see that the observation counts differ starkly from patch to patch, especially near the ends, which means that many observations were rejected due to appearing to move ``backwards'' into a patch. The $p$-values and visual inspection of the CDF plots both suggest that the Erlang distribution has a good fit for some patches, but worse for others. The fit for the terminus patches (patches 1 and 7) is especially lacking, but for some of the patches in the east-west direction (in particular patches 8, 10, and 11) the $p$-value is below $5\%$. 

\setlength{\tabcolsep}{6pt} 
\begin{table}[!ht]
	\centering
	\adjustbox{max width=\linewidth}{
		\begin{tabular}{c|cccccccccccc}
			Patch & 1 \clbox{red} & 2 \clbox{green} & 3 \clbox{blue} & 4 \clbox{yellow} & 5 \clbox{magenta} & 6 \clbox{cyan} & 7 \clbox{darkorange} & 8 \clbox{olivedrab1} & 9 \clbox{salmon} & 10 \clbox{lightskyblue2} & 11 \clbox{khaki3} & 12 \clbox{lavenderblush3}\\\hline
N. obs. &  30 & 166 & 193 & 201 & 183 & 174 & 60 & 103 & 111 & 105 & 100 & 21 \\ 
 AD &  2.6455 & 1.6557 & 0.759 & 0.853 & 0.86 & 1.4794 & 7.5402 & 2.8301 & 1.2526 & 4.3316 & 3.3847 & 0.6046 \\ 
$p$-value &  0.042 & 0.1434 & 0.5114 & 0.4442 & 0.4395 & 0.1815 & 2e-04 & 0.0335 & 0.2483 & 0.006 & 0.0176 & 0.6416 \\ 
		\end{tabular}
	}
	\caption{Same as Table~\ref{tab: airlink gof}, but for the 12 patches of the Bellevue Express.}
	\label{tab: seattle gof}
\end{table}
\setlength{\tabcolsep}{6pt} 

\begin{figure}[!th]
\begin{center}
\adjustbox{max width=\linewidth}{
	\begin{tabular}{ccc}
		\subfloat[]{\includegraphics[width=0.45\textwidth]{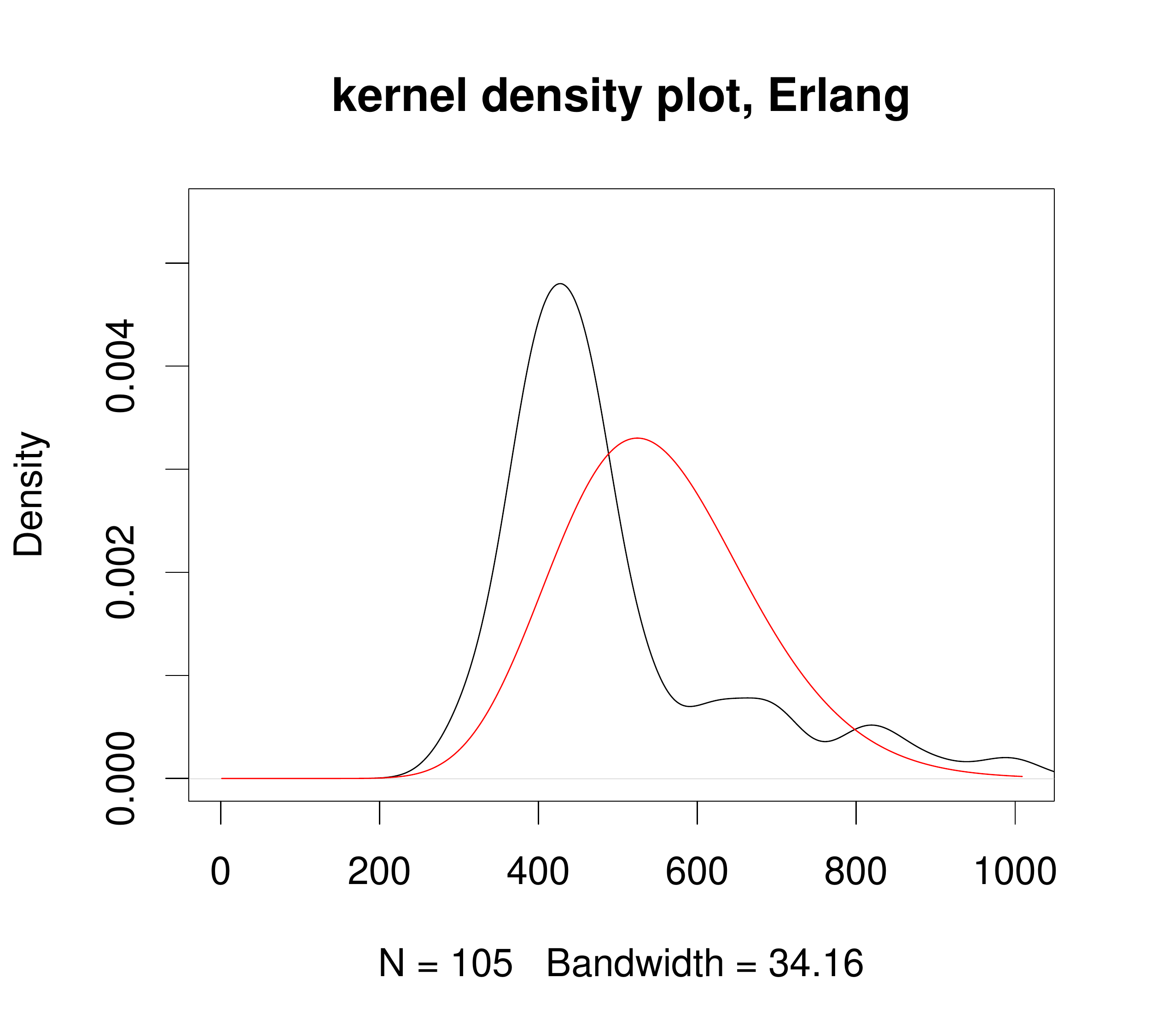} \label{fig: kerndens erlang}} 
		& \subfloat[]{\includegraphics[width=0.45\textwidth]{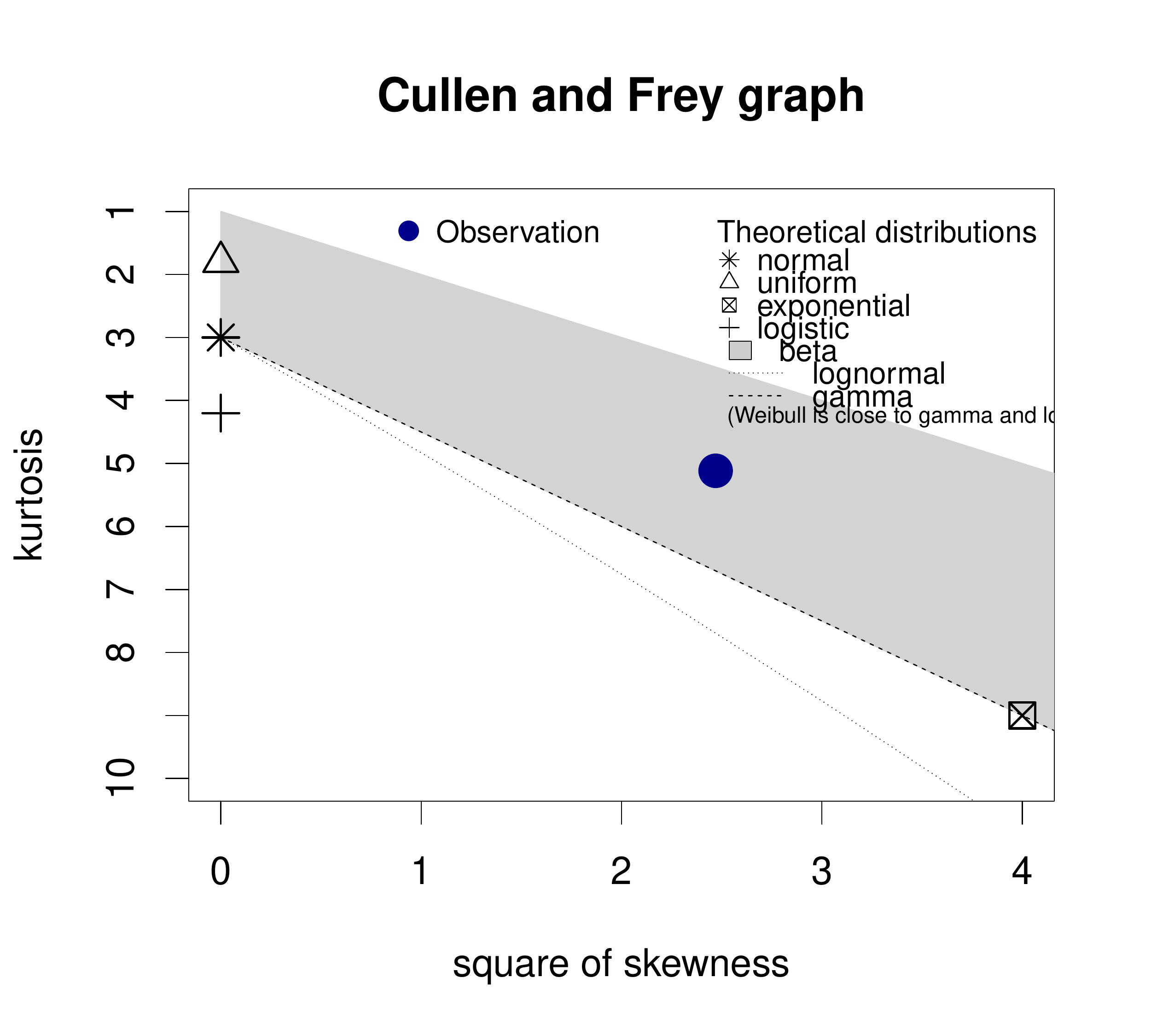} \label{fig: cullen frey}}
		& \subfloat[]{\includegraphics[width=0.45\textwidth]{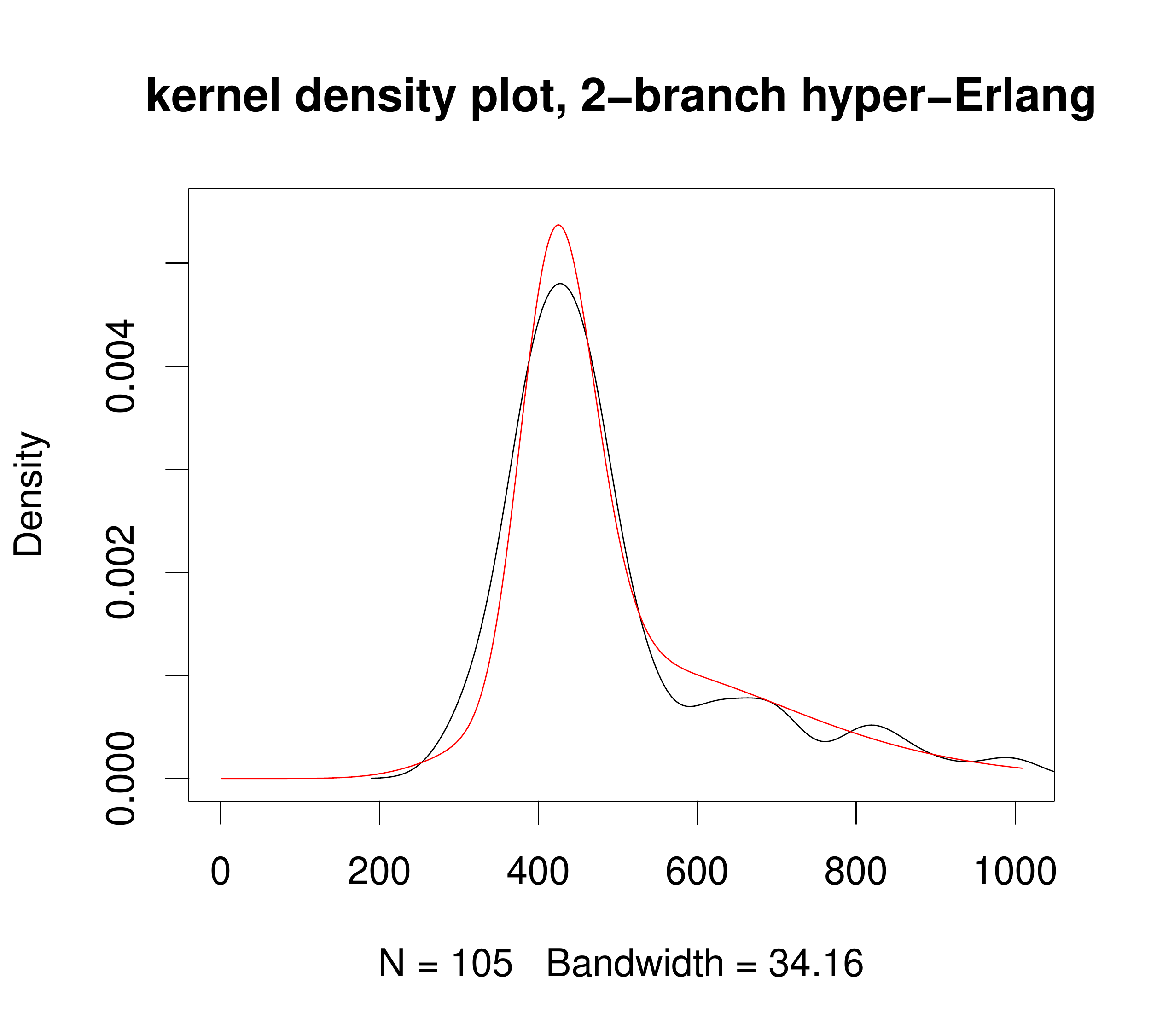} \label{fig: kerndens herlang}}
	\end{tabular}
	}
\end{center}
\caption{Comparisons between a kernel density plot of the patch crossing time observations of Patch 10 in black with a fitted Erlang distribution (Figure~\ref{fig: kerndens erlang}) and a fitted 2-branch hyper-Erlang distribution (Figure~\ref{fig: kerndens herlang}) in red. Figure~\ref{fig: cullen frey} displays a Cullen and Frey graph, which compares the skewness and kurtosis of the observations to several commonly-used distributions.}
\label{fig: gof investigation}
\end{figure}

In Figure~\ref{fig: gof investigation}, we take a closer look at the goodness-of-fit for the patch crossing time observations for Patch 10, which in Table~\ref{tab: seattle gof} can be seen to have the worst fitting results apart from the terminus patches. In Figure~\ref{fig: kerndens erlang}, we have displayed both a kernel density plot for the data and the pdf of the fitted Erlang distribution. It can be seen that the data has a considerably heavier tail on the right than the data. This is confirmed by the Cullen and Frey graph of Figure~\ref{fig: cullen frey}, from which we observe that the skewness of the data is much higher what we would expect from a Gamma distribution (which includes the Erlang) with the same kurtosis. It also tells us that other common heavy-tailed distributions, such as the lognormal and Weibull distributions, also do not have a good fit in terms of skewness and kurtosis. In Figure~\ref{fig: kerndens herlang}, we have compared the data to a 2-branch hyper-Erlang distributions. The parameters  were obtained using HyperStar, and equal $\vec{k} = (10,84)$, $\vec{\lambda} \approx (0.0171, 0.1961)$, and $\vec{\alpha} \approx (0.4762, 0.5238)$. It can be seen to have a much better fit to the data: after drawing a random sample of size $10\,000$ from this hyper-Erlang distribution using the \texttt{mapfit} library of R, we conducted a 2-sample Anderson-Darling test to compare this sample to the observation data. The resulting $p$-value (0.5551) suggests that this distribution has a much better fit to the data. However, there may be a risk of overfitting due to the high number of parameters (namely 6) in a 2-branch hyper-Erlang distribution.

\setlength{\tabcolsep}{-7pt} 
\setlength{\tabcolsep}{6pt} 

\subsubsection*{Performance Evaluation}

The statistical model checking results for the Bellevue Express are displayed in Table~\ref{tab: seattle performance}. Table~\ref{tab: seattle with timetable} presents the results for the standard setting, \emph{i.e.}, with an implicit timetable. The timetable is derived as follows: we first compute average completion time of the complete route (namely 6323 seconds, which can be obtained by summing all the $\mu$-values in Table~\ref{tab: seattle patch parameters}). We then divide this value by 12, as around 12 buses are typically observed to do the route during the rush hour period, leading to scheduled inter-departure time of 527 seconds. However, in our experiments we found that the timetabled setting is overly optimistic. For example, by computing empirical EWTs by recording the observed patch departure times in our dataset (which can be noisy because dropped observations result in a bias towards high values) and applying $\text{EWT} = \mu/(2\sigma^2)$, we found that the empirical EWTs for patches 2, 3, 4, 5, and 6 were approximately equal to $204.6$, $183.0$, $172.8$, $188.0$, and $216.6$ respectively. Not only are these values much higher than the values in Table~\ref{tab: seattle with timetable}, they do not exhibit the expected behaviour of the EWT being higher in patches that are further away from the previous terminus. Hence, in Table~\ref{tab: seattle without timetable} we also display the result for the situation where there is no headway correction at all. In this setting, in steady-state all three performance measures are the same across the patches. The EWT and EVWT requirements are not met, being comfortably above 75 seconds and 5$\%$ respectively, although the BPH requirement is still met. Interestingly, the EVWT in patch 12 is higher in the situation with a timetable than without, which means that such a strategy is not necessarily helpful in all patches along the route. 

\setlength{\tabcolsep}{0pt} 
\begin{table}
\makebox[\textwidth][c]{
\subfloat[With a timetable at the end of the patches]{
	\begin{tabular}{c|cccccccccccc}
			Patch & 1 \clbox{red} & 2 \clbox{green} & 3 \clbox{blue} & 4 \clbox{yellow} & 5 \clbox{magenta} & 6 \clbox{cyan} & 7 \clbox{darkorange} & 8 \clbox{olivedrab1} & 9 \clbox{salmon} & 10 \clbox{lightskyblue2} & 11 \clbox{khaki3} & 12 \clbox{lavenderblush3}\\\hline
			EWT\; & \perrthin{0.52}{0.02} & \perrthin{7.61}{0.19} & \perrthin{10.31}{0.29} & \perrthin{15.30}{0.43} & \perrthin{27.23}{0.76} & \perrthin{41.78}{0.90} & \perrthin{0.22}{0.02} & \perrthin{23.78}{0.65} & \perrthin{48.74}{1.13} & \perrthin{71.82}{1.69} & \perrthin{82.62}{1.62} & \perrthin{93.32}{1.30}\\
			EVWT\; & ---  & \perrthin{0.000}{0.00} & \perrthin{0.000}{0.00} & \perrthin{0.002}{0.00} & \perrthin{0.014}{0.00} & \perrthin{0.037}{0.00} & ---  & \perrthin{0.010}{0.00} & \perrthin{0.053}{0.00} & \perrthin{0.096}{0.01} & \perrthin{0.111}{0.01} & \perrthin{0.129}{0.01}\\
			BPH\; & \perrthin{0.000}{0.00} & ---  & \perrthin{0.000}{0.00} & \perrthin{0.000}{0.00} & \perrthin{0.001}{0.00} & \perrthin{0.003}{0.00} & ---  & \perrthin{0.000}{0.00} & \perrthin{0.005}{0.00} & \perrthin{0.012}{0.00} & \perrthin{0.020}{0.00} & \perrthin{0.028}{0.00}
		\end{tabular}
			\label{tab: seattle with timetable}
	}
	}
	
	\vspace{0.1cm}
	
		\makebox[\textwidth][c]{
		\subfloat[Without a timetable at the end of the patches]{
		\begin{tabular}{c|cccccccccccc}
			Patch & 1 \clbox{red} & 2 \clbox{green} & 3 \clbox{blue} & 4 \clbox{yellow} & 5 \clbox{magenta} & 6 \clbox{cyan} & 7 \clbox{darkorange} & 8 \clbox{olivedrab1} & 9 \clbox{salmon} & 10 \clbox{lightskyblue2} & 11 \clbox{khaki3} & 12 \clbox{lavenderblush3}\\\hline
			EWT\; & \perrthin{152.07}{8.00} & \perrthin{147.23}{5.67} & \perrthin{157.83}{7.76} & \perrthin{145.14}{7.48} & \perrthin{155.21}{8.31} & \perrthin{151.79}{7.24} & \perrthin{150.60}{10.86} & \perrthin{151.65}{10.75} & \perrthin{153.24}{9.02} & \perrthin{159.42}{12.43} & \perrthin{151.26}{7.63} & \perrthin{145.47}{7.51}\\
			EVWT\; & \perrthin{0.112}{0.01} & \perrthin{0.113}{0.01} & \perrthin{0.117}{0.01} & \perrthin{0.110}{0.01} & \perrthin{0.121}{0.01} & \perrthin{0.110}{0.01} & \perrthin{0.113}{0.01} & \perrthin{0.128}{0.01} & \perrthin{0.114}{0.01} & \perrthin{0.114}{0.00} & \perrthin{0.116}{0.01} & \perrthin{0.114}{0.01}\\
			BPH\; & \perrthin{0.026}{0.01} & \perrthin{0.028}{0.00} & \perrthin{0.029}{0.01} & \perrthin{0.032}{0.01} & \perrthin{0.024}{0.01} & \perrthin{0.030}{0.00} & \perrthin{0.026}{0.01} & \perrthin{0.026}{0.00} & \perrthin{0.025}{0.01} & \perrthin{0.026}{0.01} & \perrthin{0.028}{0.01} & \perrthin{0.027}{0.00}
		\end{tabular}
		\label{tab: seattle without timetable}
		}
		
}
\vspace{0.1cm}

	\caption{Same as Table~\ref{tab: patch parameters}, but for the 12 patches of the Bellevue Express in two settings: with a timetable at both termini (as described in Section~\ref{sec: model checking}), and without.}
	\label{tab: seattle performance}
\end{table}
\setlength{\tabcolsep}{6pt} 

\subsection{Service Improvement Strategies for the Airlink} \label{sec:strategies}

As mentioned previously, the simulation model can be used to investigate the impact of hypothetical changes to a service on its performance. In Table~\ref{tab: airlink simulation results}, we have presented an overview of the simulation results for different headway correction strategies. We first include the EWT as computed from the data. Next is the situation where the services wait at the termini using the procedure described in Section~\ref{sec: model checking}. Note that the difference between the empirical EWT and its simulated counterpart is greater in the east-west direction (as evidenced by the EWT in Patch 10) than in the reverse. The next row corresponds to the situation where there is no headway correction \emph{even} at the termini. With buses no longer standing still at the termini, the headway variance is much larger than in the standard setting. However, the mean observed headway ($\hat{\mu}_Y$) is 150 seconds lower, which is a substantial improvement. Because the EWT is defined as the observed average waiting minus the timetabled waiting time, we subtract the \emph{originally} timetabled waiting time. This leads to a net effect of -15 seconds in all patches, compared to between 3 and 32 seconds in the original setting. This means that the increased headway variance is outweighed by the gains in average headway.

To improve system performance further, \cite{moreira2015improving} discusses four methods to reduce headway variance: \emph{bus holding} (increasing dwell times at stops), \emph{speed modification} (decreasing maximum cruise speed), \emph{stop-skipping}, and \emph{short-turning} (switching directions before the end stop). Since the system state in our model does not distinguish whether buses are moving or standing still, only the fourth of these can be made explicit. However, we can try to capture the behaviour of speed modification (and potentially stop-skipping) in an abstract way by lowering the rate at which buses complete the patches. Also, we can mimic bus holding by lower-bounding the exit times from patches based on the headway with the previous bus.
The procedure that we have chosen is as follows. For bus holding, we impose that buses cannot leave patch $j$, $j \in \{1,\ldots,\numpatches\}$, unless the clock $y_j$, which represents the time since the previous bus departure from patch~$j$, is greater than or equal to some threshold value $\theta$. For speed modification, at each time step, we compute for each bus its route progression using the procedure discussed at the end of Section~\ref{sec: map generation}. With 11 buses on the road, the difference in terms of route progression between subsequent buses should be roughly $9.09\%$ (this can vary quite a bit near the end points). We then introduce the following mechanism: if the route progression difference between bus $i$ and the one following it is greater than $100\% \cdot \theta$, we slow bus $i$ down (via its phase completion rate) by $10\%$. 
After all, in order to maintain headway regularity it may be preferable to use the possibility to go slightly slower than specified by the speed limit to avoid getting too close to the bus in front.
This can be combined with timetables at the termini.

The results displayed in Table~\ref{tab: airlink simulation results} indicate that bus holding combined with the timetable does have a stabilising effect, with a slightly higher EWT in the patches directly after the termini (presumably due to a higher probability of a bus taking so long to complete the route that it has run out of ``slack'' time at the terminus), but a lower EWT in the middle of the route. If the threshold $\theta$ is set too high, then route completion times are such that buses tend to arrive at the termini after their timetabled departure time, leading to higher average headways and higher EWTs. By contrast, the impact of speed modification with the chosen parameters is very small. The best performance is achieved by a combination of bus holding with a lack of timetables at the termini, leading to an excess waiting time of between -80 and -100 seconds. Note that reducing the EWT may result in other performance metrics being affected: \emph{e.g.}, the Averaged In-Vehicle Times (AIVT) as discussed in \cite{moreira2016online}, which concerns the time spent by passengers in vehicles. Passengers are currently not part of the model, so we leave this as subject for further research.

\setlength{\tabcolsep}{-3.5pt}
\begin{table}[!b]
	\makebox[\textwidth][c]{
		\begin{tabular}{cc|cccc|cccc}
			& & \multicolumn{4}{c}{EWT per patch} & \multicolumn{4}{c}{$\hat{\mu}_Y$ per patch} \\
			&  & 2 & 5 & 7 & 10 & 2 & 5 & 7 & 10 \\ 
			\textit{Observed from the data} & & 16.20 & 23.02 & 11.35 & 41.93 & 485.21 & 474.45 & 477.19 & 495.10 \\ \hline
			\multicolumn{6}{c}{ } \\[-0.35cm]  \hline
			Simulation with strategy: & $\theta$ & & & & \\
			Timetable at end points only & & \perr{4.99}{0.05} & \perr{26.02}{0.18} & \perr{3.73}{0.01} & \perr{31.29}{0.23} & \perr{478.44}{0.04} & \perr{478.40}{0.04} & \perr{478.56}{0.05} & \perr{478.47}{0.02} \\ \hline 
			No headway correction at all & & \perr{-12.03}{3.46} & \perr{-14.16}{3.39} & \perr{-15.58}{3.43} & \perr{-15.01}{3.69} & \perr{324.42}{0.10} & \perr{324.34}{0.08} & \perr{324.47}{0.12} & \perr{324.36}{0.07} \\ \hline 
			\multirow{3}{*}{Bus holding in all patches} & \hspace{0.5cm} $120.0$ \hspace{0.5cm} & \perr{4.95}{0.04} & \perr{15.82}{0.10} & \perr{3.68}{0.01} & \perr{22.02}{0.12} & \perr{478.39}{0.04} & \perr{478.43}{0.04} & \perr{478.50}{0.04} & \perr{478.44}{0.03} \\ 
			 & $240.0$ & \perr{28.10}{0.04} & \perr{37.75}{0.13} & \perr{42.46}{0.14} & \perr{41.88}{0.14} & \perr{501.38}{0.07} & \perr{500.81}{0.09} & \perr{501.38}{0.10} & \perr{501.18}{0.10} \\ 
			 & $360.0$ & \perr{140.42}{0.05} & \perr{149.46}{0.12} & \perr{152.76}{0.11} & \perr{154.60}{0.20} & \perr{614.30}{0.10} & \perr{612.90}{0.11} & \perr{614.41}{0.14} & \perr{614.08}{0.13} \\ 
			 \hline 
			\multirow{5}{*}{Bus holding only (all patches)} & \hspace{0.5cm} $60.0$ \hspace{0.5cm} & \perr{-98.02}{0.17} & \perr{-91.00}{0.24} & \perr{-85.96}{0.20} & \perr{-85.58}{0.28} & \perr{354.32}{0.09} & \perr{354.24}{0.08} & \perr{354.32}{0.07} & \perr{354.23}{0.08} \\ 
			 & $120.0$ & \perr{-68.02}{0.08} & \perr{-58.77}{0.13} & \perr{-53.16}{0.11} & \perr{-53.45}{0.19} & \perr{399.22}{0.10} & \perr{399.12}{0.08} & \perr{399.27}{0.10} & \perr{399.19}{0.08} \\ 
			 & $180.0$ & \perr{-22.76}{0.04} & \perr{-13.02}{0.10} & \perr{-7.71}{0.11} & \perr{-8.32}{0.18} & \perr{448.82}{0.10} & \perr{448.57}{0.07} & \perr{448.85}{0.09} & \perr{448.69}{0.09} \\ 
			 & $240.0$ & \perr{28.00}{0.05} & \perr{37.64}{0.12} & \perr{42.39}{0.10} & \perr{42.00}{0.20} & \perr{501.30}{0.11} & \perr{500.79}{0.09} & \perr{501.38}{0.10} & \perr{501.11}{0.09} \\ 
			 & $300.0$ & \perr{82.57}{0.03} & \perr{92.12}{0.12} & \perr{95.91}{0.09} & \perr{96.39}{0.19} & \perr{556.44}{0.11} & \perr{555.68}{0.08} & \perr{556.59}{0.12} & \perr{556.29}{0.10} \\ 
			 \hline 
			\multirow{3}{*}{Speed modification} & $0.05$ & \perr{4.94}{0.05} & \perr{24.01}{0.15} & \perr{3.70}{0.01} & \perr{34.41}{0.21} & \perr{478.44}{0.03} & \perr{478.41}{0.03} & \perr{478.56}{0.05} & \perr{478.45}{0.04} \\ 
			 & $0.01$ & \perr{4.96}{0.04} & \perr{25.80}{0.19} & \perr{3.70}{0.01} & \perr{41.41}{0.28} & \perr{478.44}{0.03} & \perr{478.40}{0.02} & \perr{478.57}{0.03} & \perr{478.45}{0.04} \\ 
			 & $0.15$ & \perr{5.23}{0.05} & \perr{30.56}{0.25} & \perr{3.68}{0.02} & \perr{37.21}{0.24} & \perr{478.54}{0.03} & \perr{478.39}{0.04} & \perr{478.54}{0.05} & \perr{478.43}{0.04} \\ 
		\end{tabular}
	}

\caption{Comparison of the impact of several performance improvement strategies in terms of their EWT and observed average headway $\hat{\mu}_Y$.}
	\label{tab: airlink simulation results}
\end{table}
\setlength{\tabcolsep}{6.0pt}


\section{Conclusions \& Discussion} \label{sec: conclusions}

In this paper, we have presented a novel, fully automated approach for using AVL data to build and parameterise stochastic models of transport services. The model can be used to obtain reliable estimates of the performance of the service, which can help operators determine whether they are meeting the requirements set by regulators. Furthermore, it can be used to analyse the impact of changes to the system. The data and the code used to conduct the experiments are available for public use.

One direction for future work is to expand the model to incorporate other real-world phenomena. For example, the current model is time-homogeneous and focuses on a specific part of the day (between 10AM to 3PM). An extension would be a fully time-dependent model including the rush hours and the night. Time-inhomogeneity is straightforwardly added to the model by assigning to each patch multiple sets of Erlang parameters, such that only one set of parameters can be chosen based on the time of the day. This can be further extended to include weekday/weekend effects and perhaps even seasonal effects. Another addition would be to incorporate weather effects, as bus systems are used more intensively during periods of heavy precipitation. To realistically parameterise such a model, the AVL dataset would need to be matched with a dataset of weather measurements.

A powerful addition to the model would be to include passengers boarding or alighting from the vehicles. However, this requires the AVL measurements to be matched with  Automatic Passenger Counting (APC) data, which can be challenging to obtain, either because this data is not being collected or because operators are understandably hesitant to share it. From the point of view of system reliability, another interesting addition would be to include vehicle break-downs, as these can have a big impact on certain performance measures (particularly the EVWT). Extending the model to include performance measures for non-frequent services, particularly the On-Time Performance (OTP), is straightforward. Finally, instead of modelling a single service, one could create a model of the full network and include contention between buses for access to bus stops.


\section*{Acknowledgements}

This work has been supported by the EU project QUANTICOL,
600708. The authors thank Bill Johnston and Philip
Lock of Lothian Buses for providing access to the data and
for their helpful feedback.

\bibliographystyle{plain}
\bibliography{biblio}

\appendix
\section{Appendix} \label{sec: appendix}

\setlength{\tabcolsep}{0pt}
\renewcommand{\arraystretch}{0}
\newcommand{\cdfscale}{0.33\textwidth}
\begin{figure}[!t]
	\begin{center}
		\makebox[\textwidth][c]{
			\begin{tabular}{ccc}
			  \subfloat{\includegraphics[width=\cdfscale]{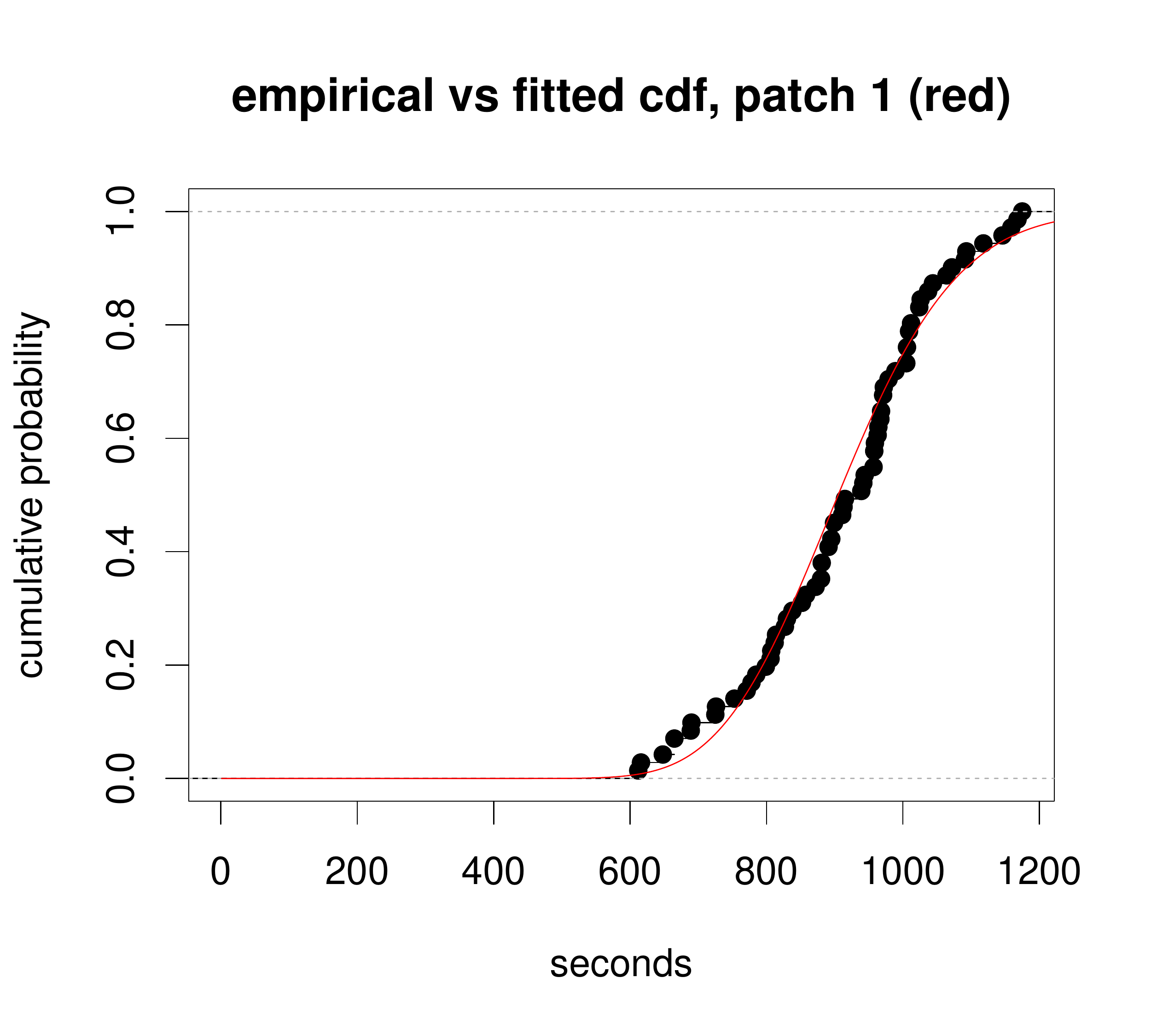} \label{fig: cdf0}}
			 & \subfloat{\includegraphics[width=\cdfscale]{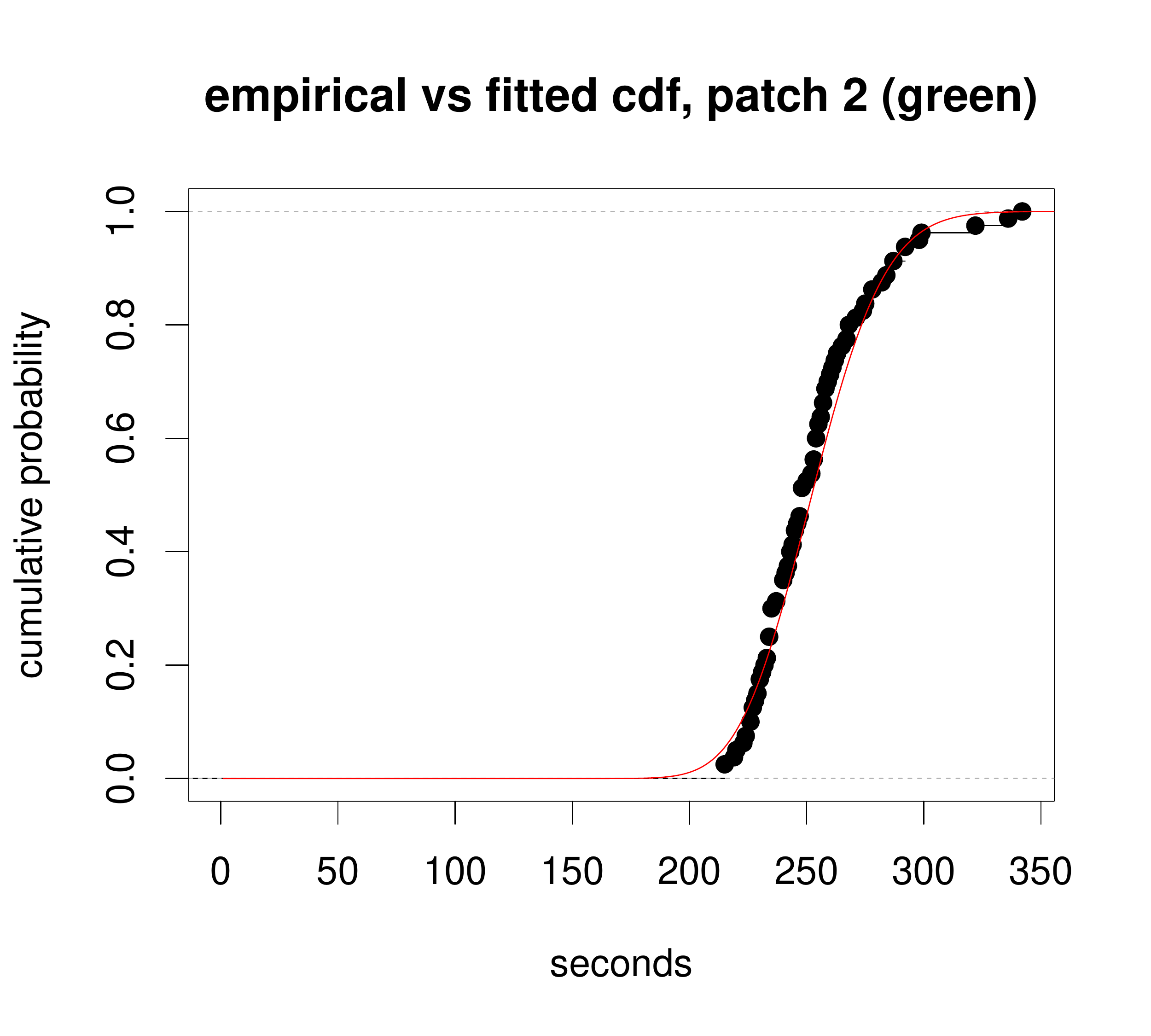} \label{fig: cdf1}}
			 & \subfloat{\includegraphics[width=\cdfscale]{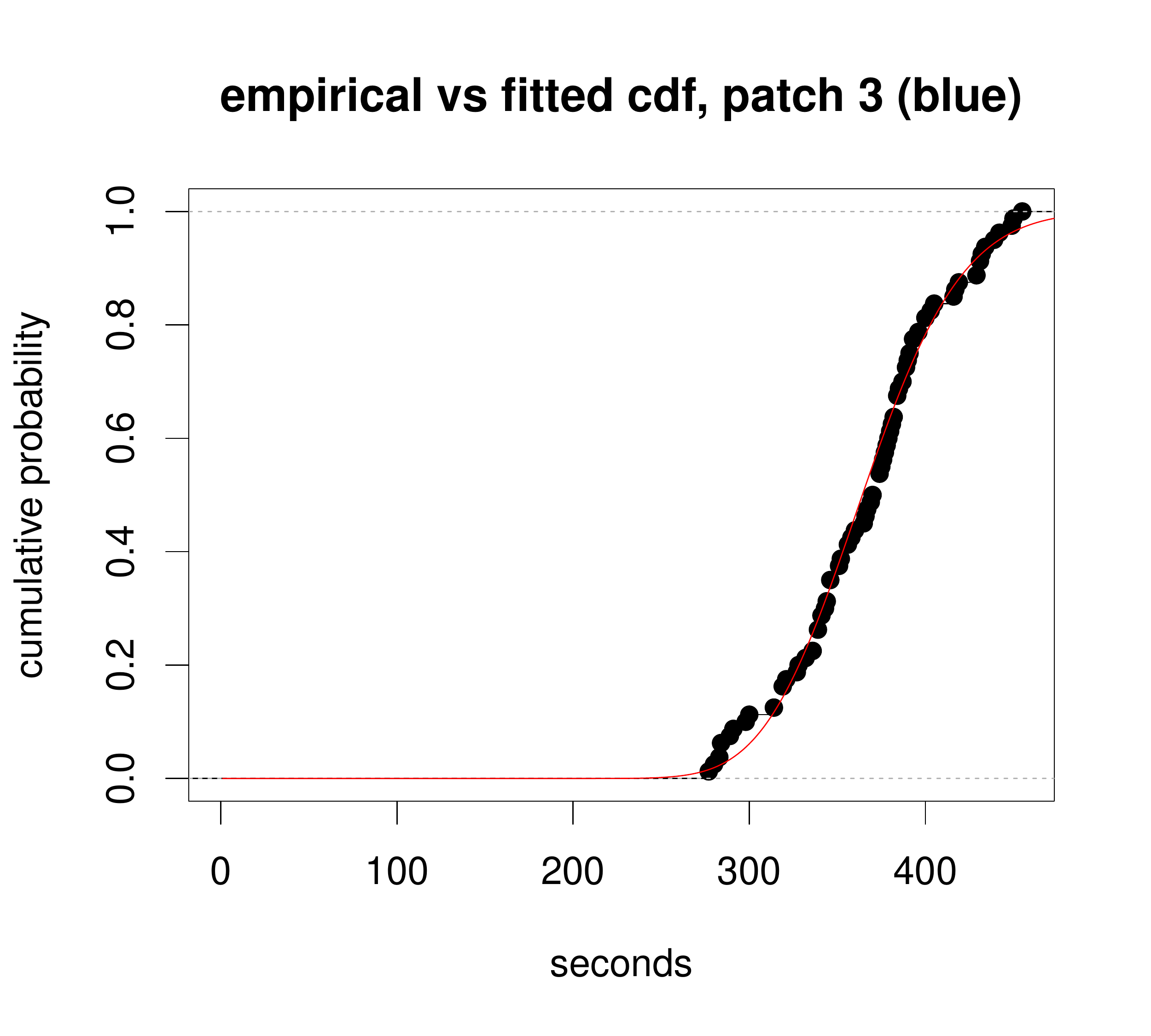} \label{fig: cdf2}} \vspace{-0.5cm} \\
			 	\subfloat{\includegraphics[width=\cdfscale]{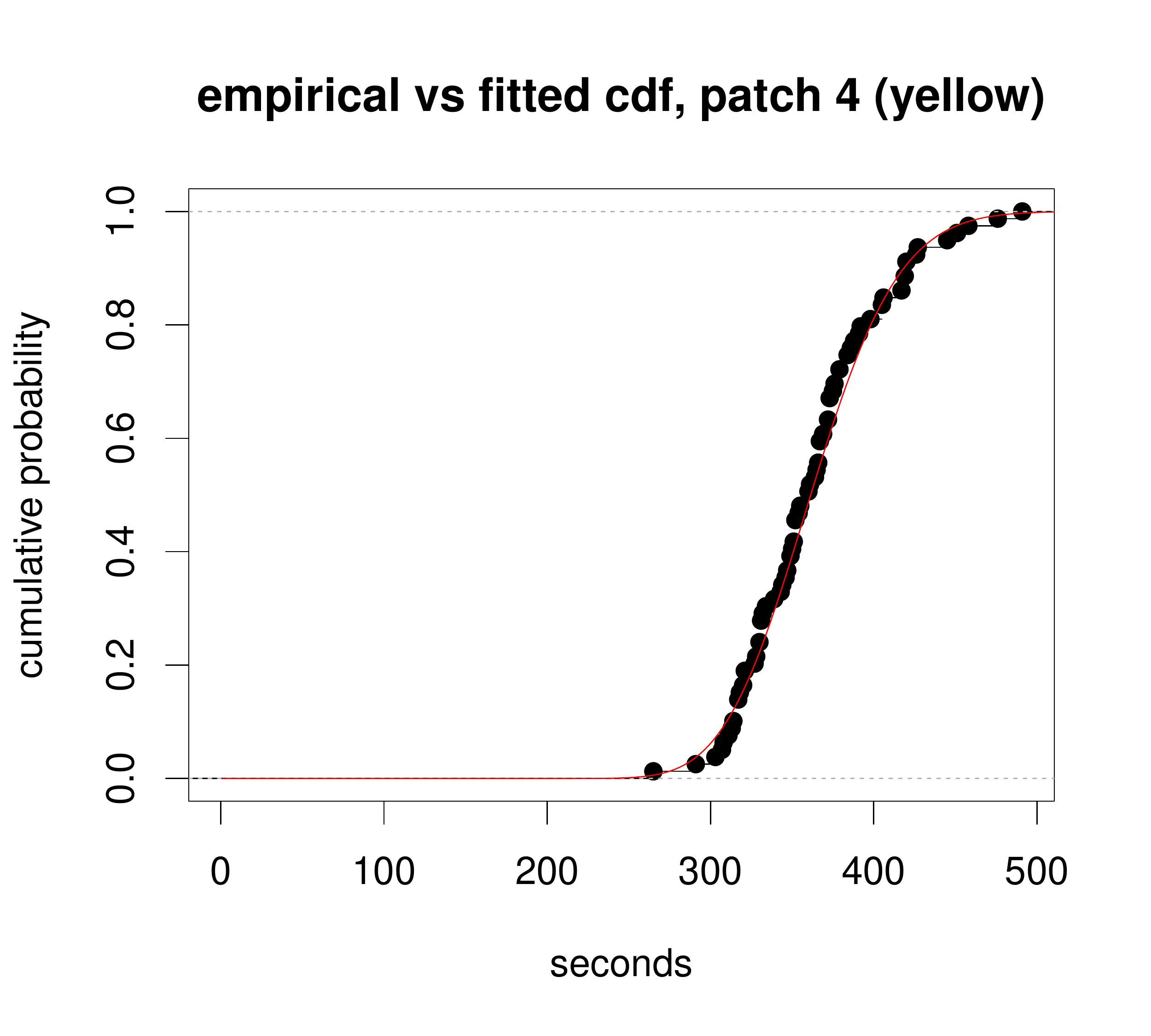} \label{fig: cdf3}}
			 & \subfloat{\includegraphics[width=\cdfscale]{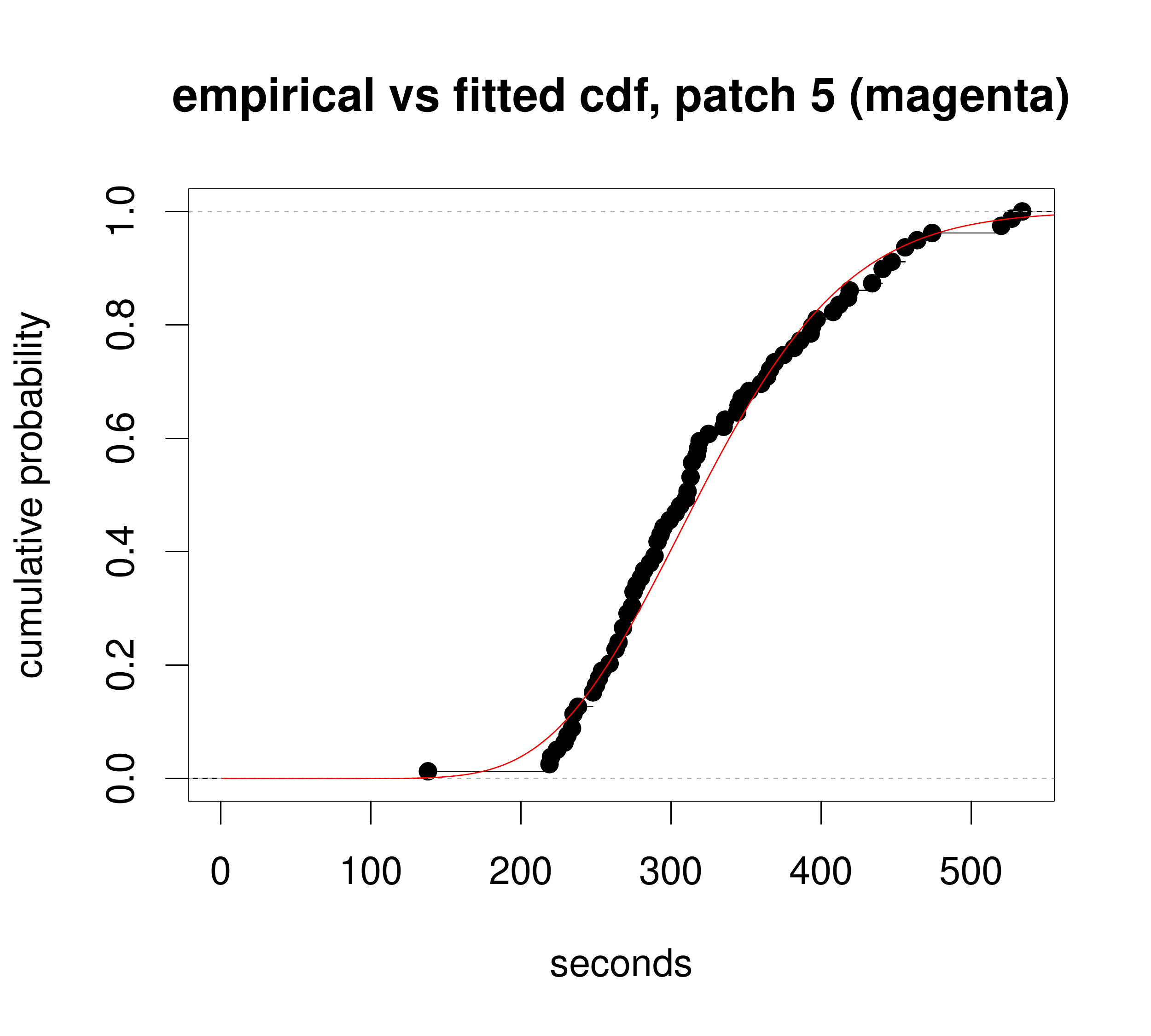} \label{fig: cdf4}}
			 & \subfloat{\includegraphics[width=\cdfscale]{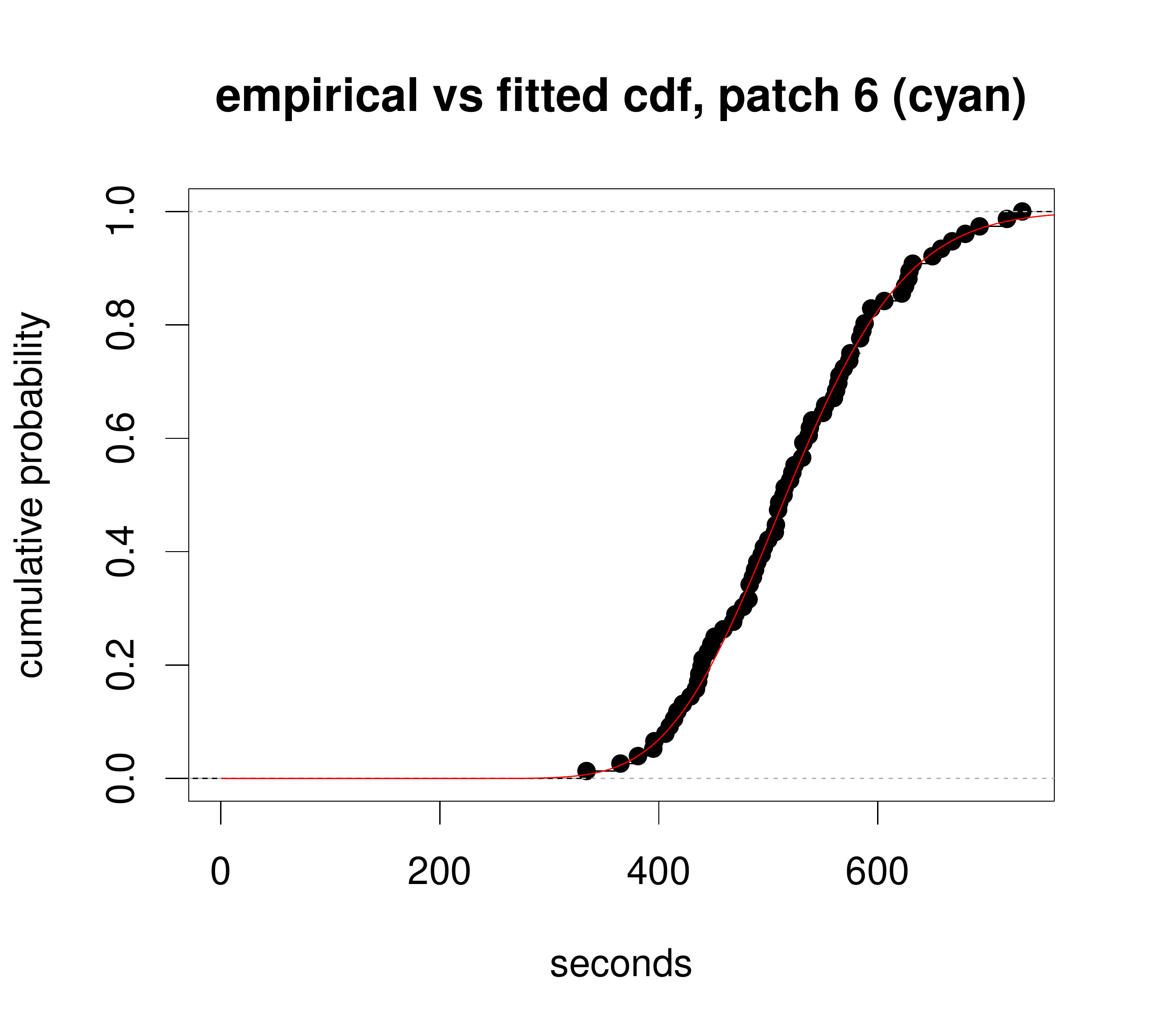} \label{fig: cdf5}}  \vspace{-0.5cm} \\
			 	\subfloat{\includegraphics[width=\cdfscale]{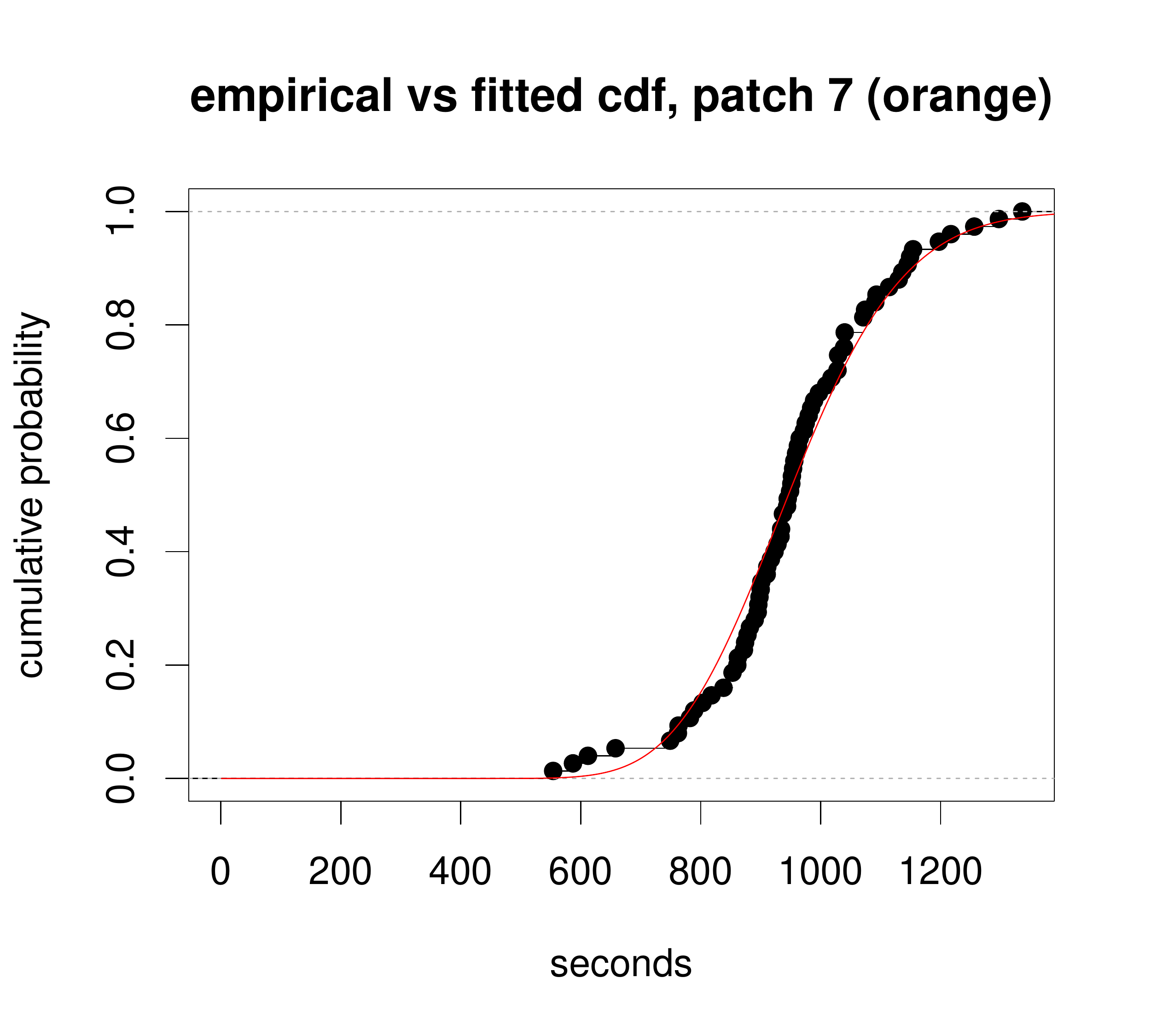} \label{fig: cdf6}} 
			 & \subfloat{\includegraphics[width=\cdfscale]{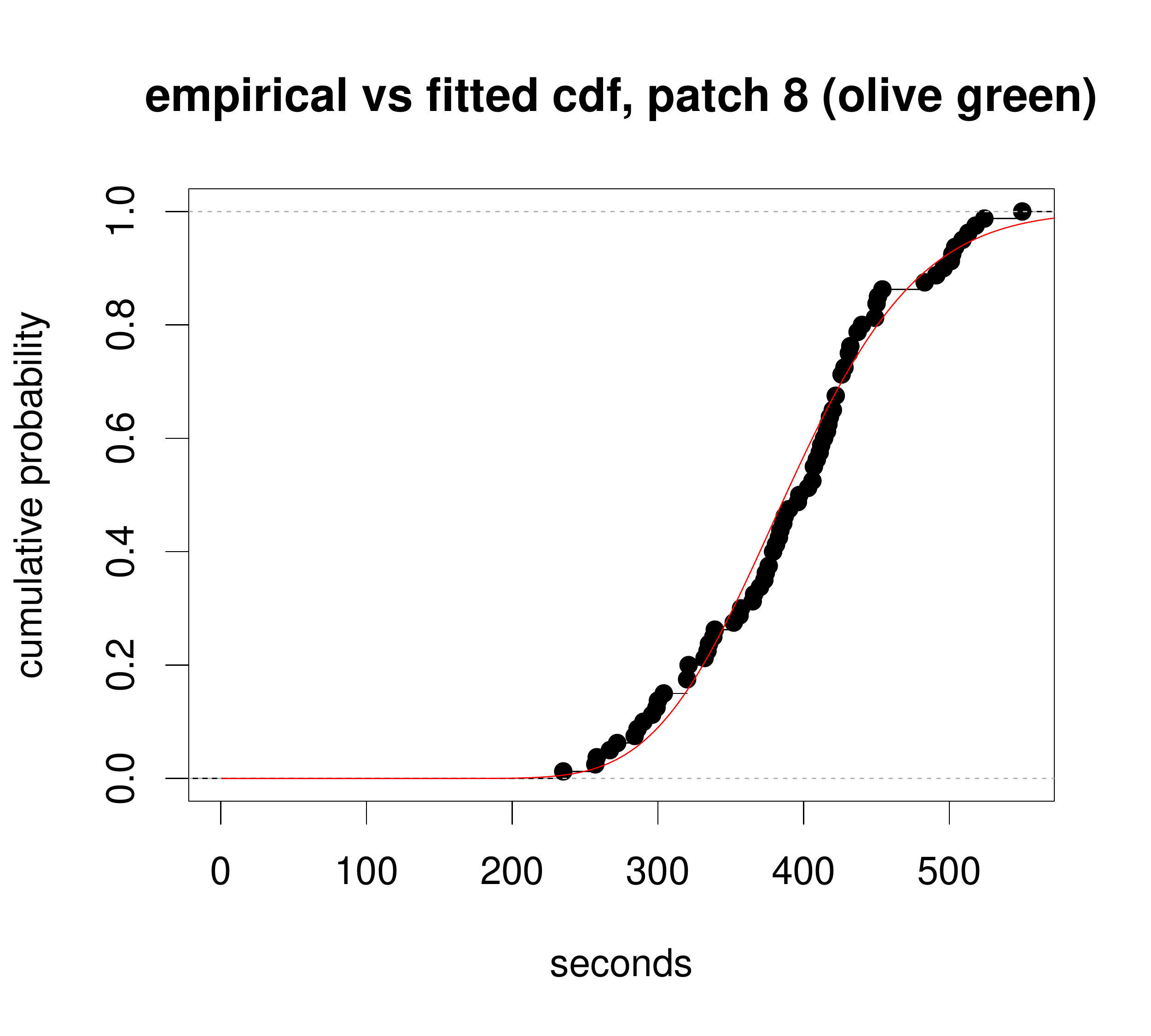} \label{fig: cdf7}}
			 & \subfloat{\includegraphics[width=\cdfscale]{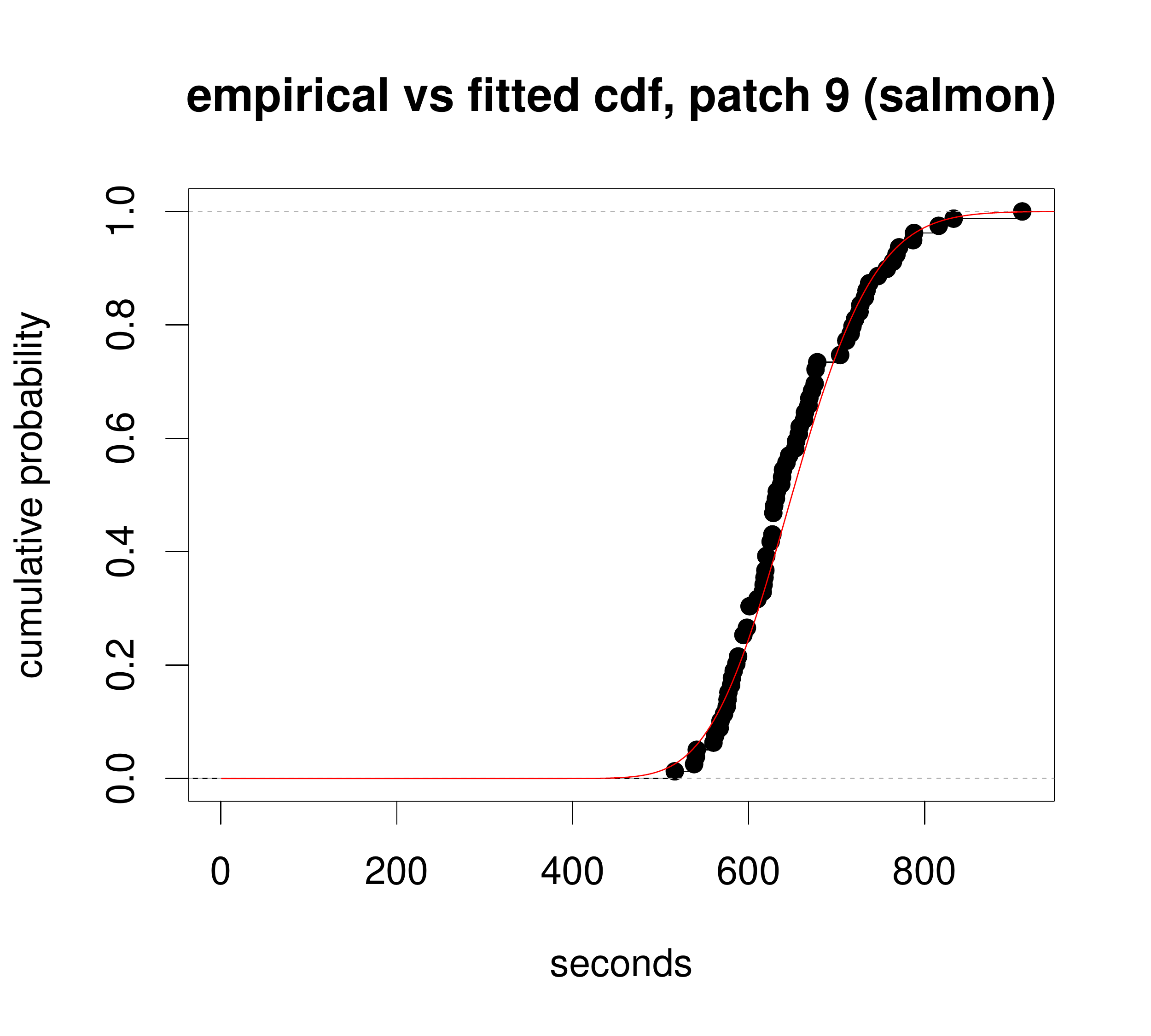} \label{fig: cdf8}}  \vspace{-0.5cm} \\
			 & \subfloat{\includegraphics[width=\cdfscale]{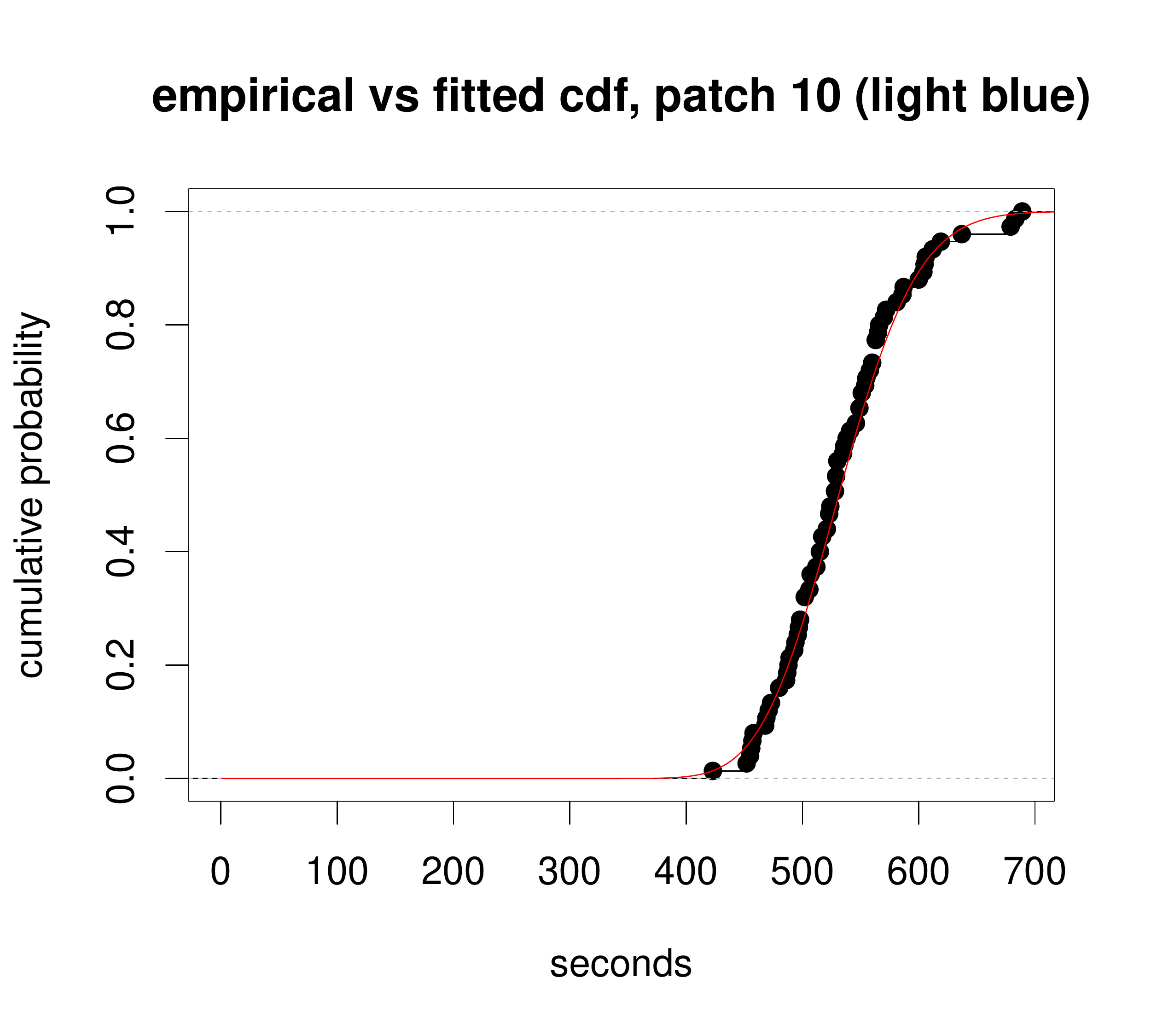} \label{fig: cdf9}}
		
			\end{tabular}
		}
	\end{center}
	\caption{Empirical (black dots) vs fitted (smooth lines) CDF plots for all of the 10 final patches in the Edinburgh dataset.}
	\label{fig: cdf plots}
\end{figure}

\newpage

\begin{figure}[!t]
	\begin{center}
		\makebox[\textwidth][c]{
			\begin{tabular}{ccc}
			  \subfloat{\includegraphics[width=\cdfscale]{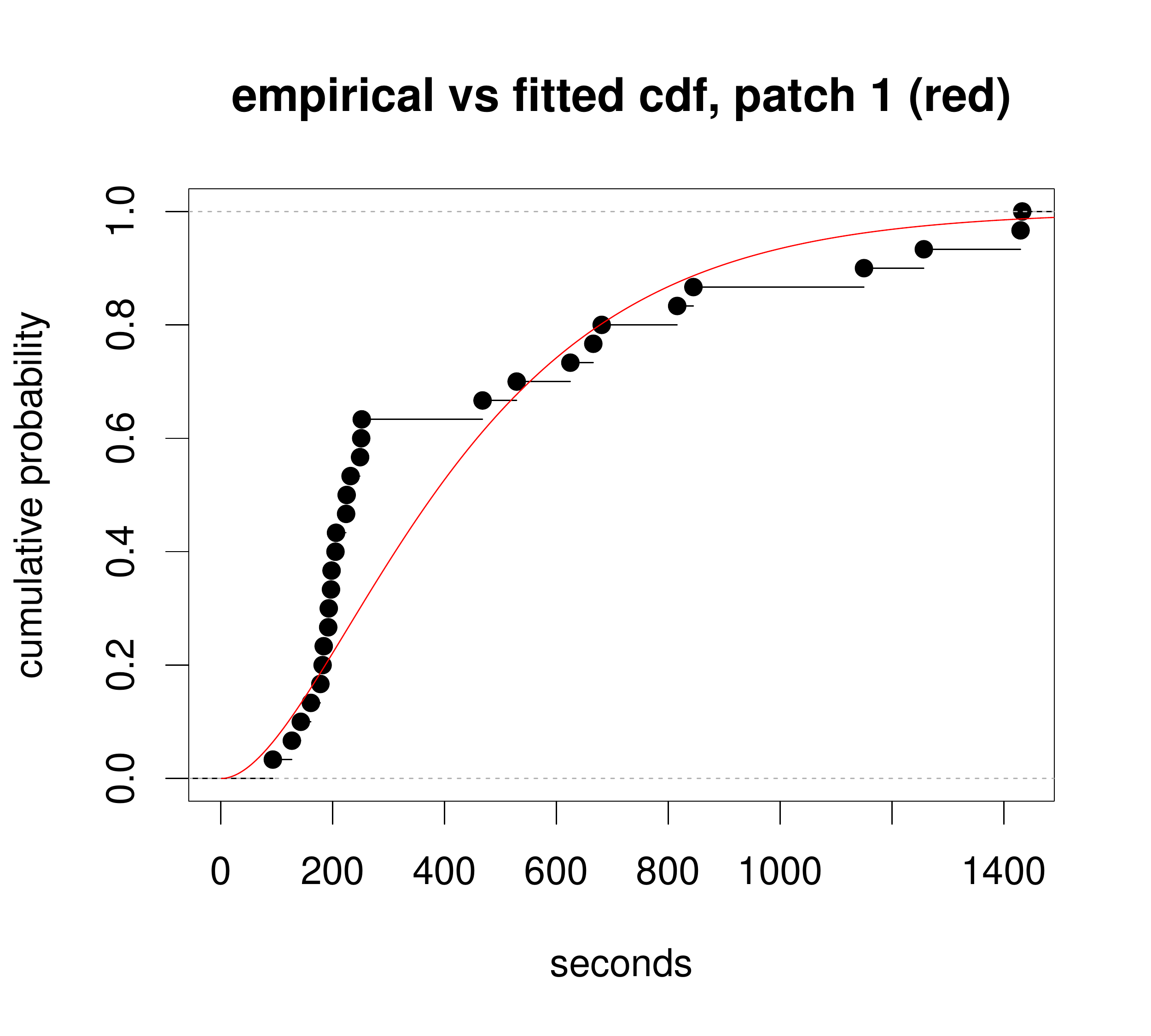} \label{fig: seattle_cdf0}}
			 & \subfloat{\includegraphics[width=\cdfscale]{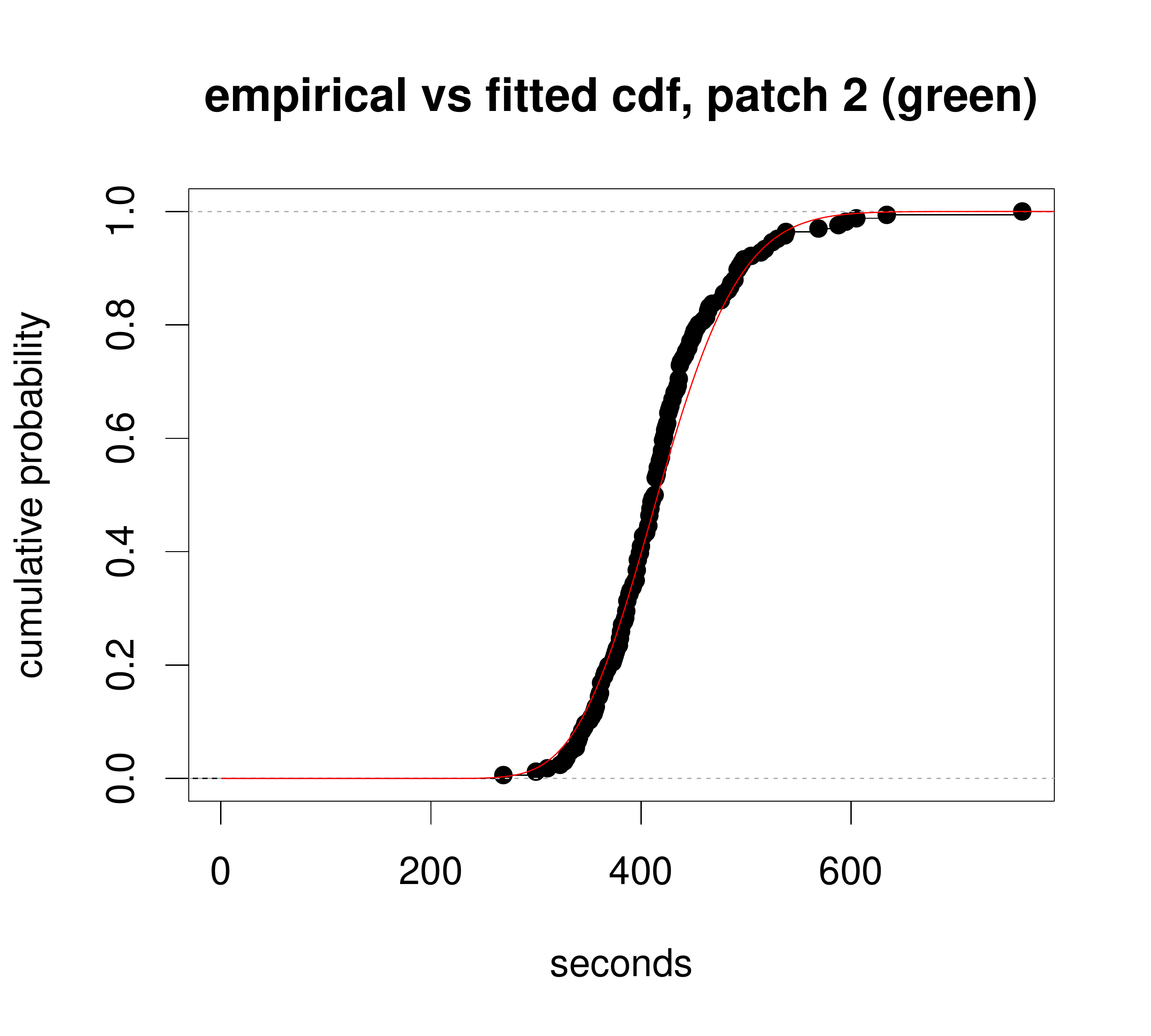} \label{fig: seattle_cdf1}}
			 & \subfloat{\includegraphics[width=\cdfscale]{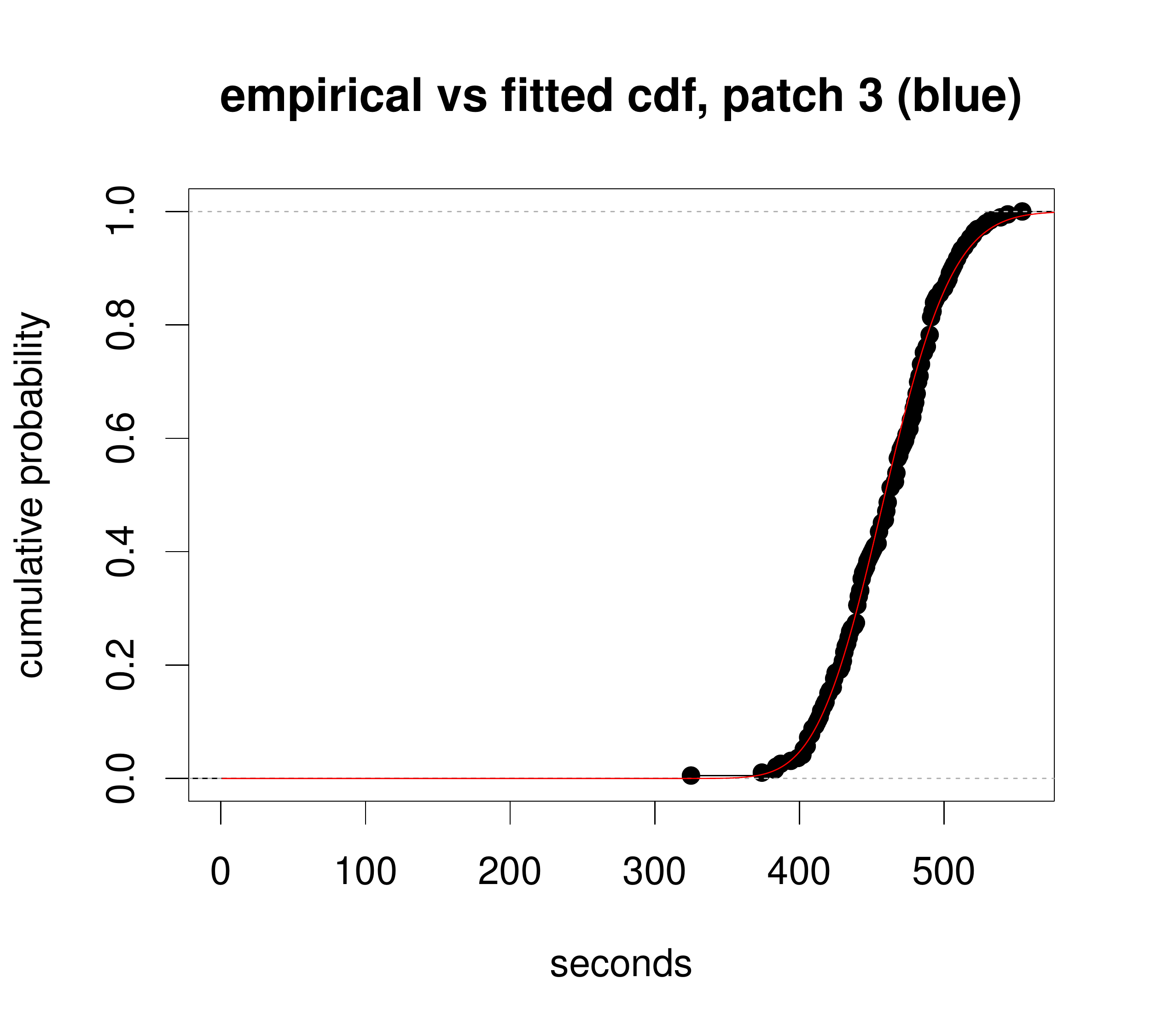} \label{fig: seattle_cdf2}} \vspace{-0.5cm} \\
			 	\subfloat{\includegraphics[width=\cdfscale]{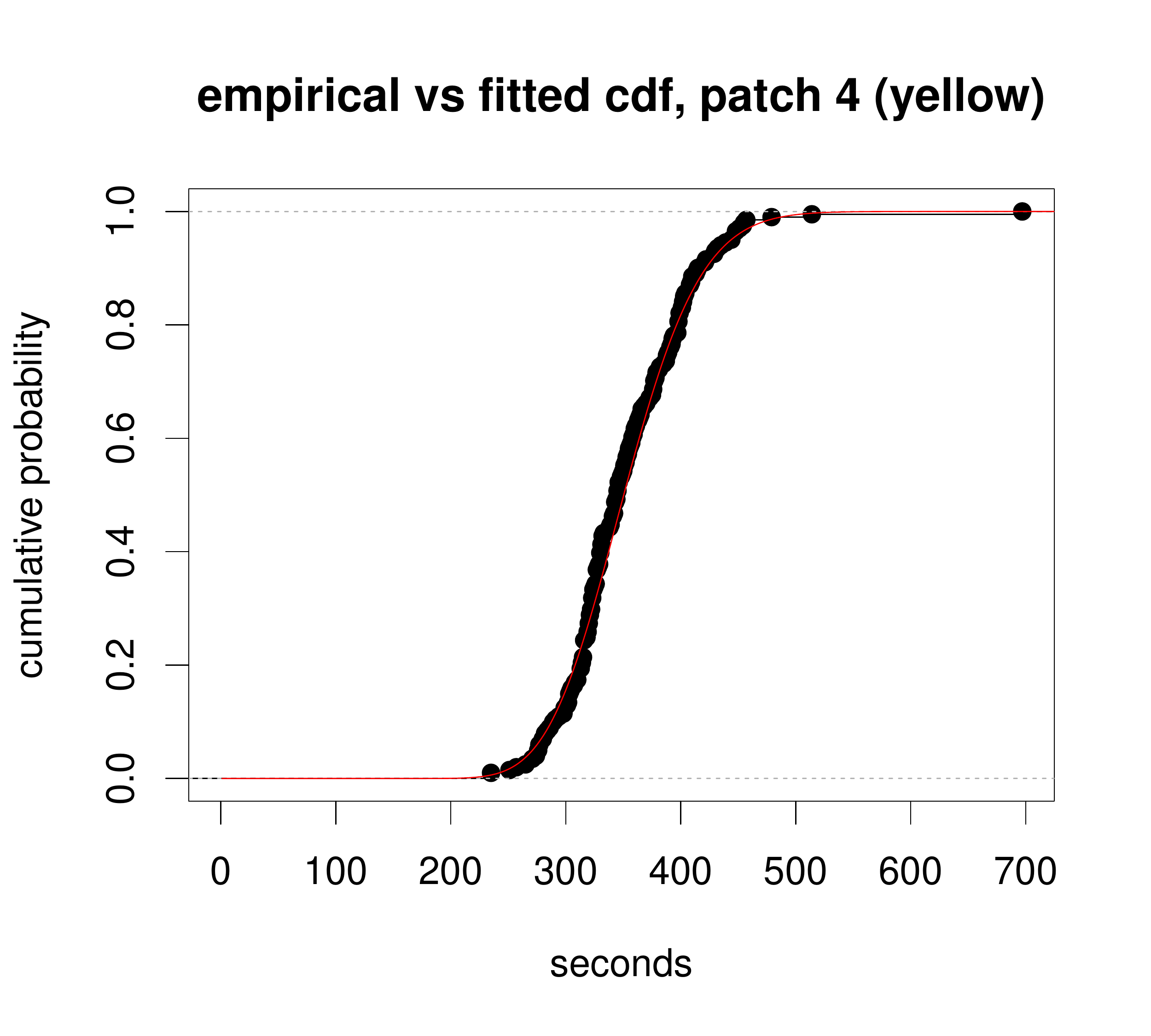} \label{fig: seattle_cdf3}}
			 & \subfloat{\includegraphics[width=\cdfscale]{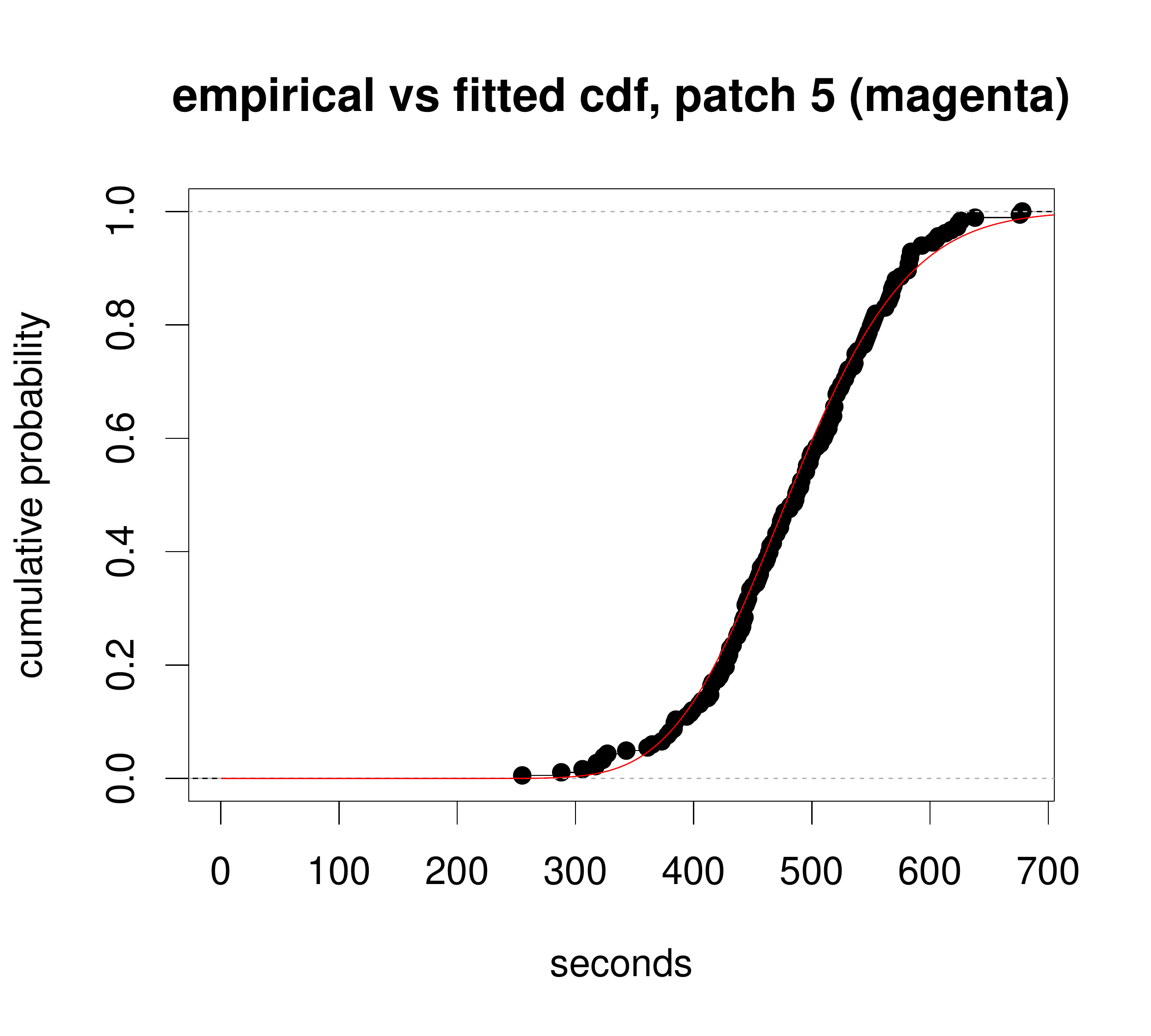} \label{fig: seattle_cdf4}}
			 & \subfloat{\includegraphics[width=\cdfscale]{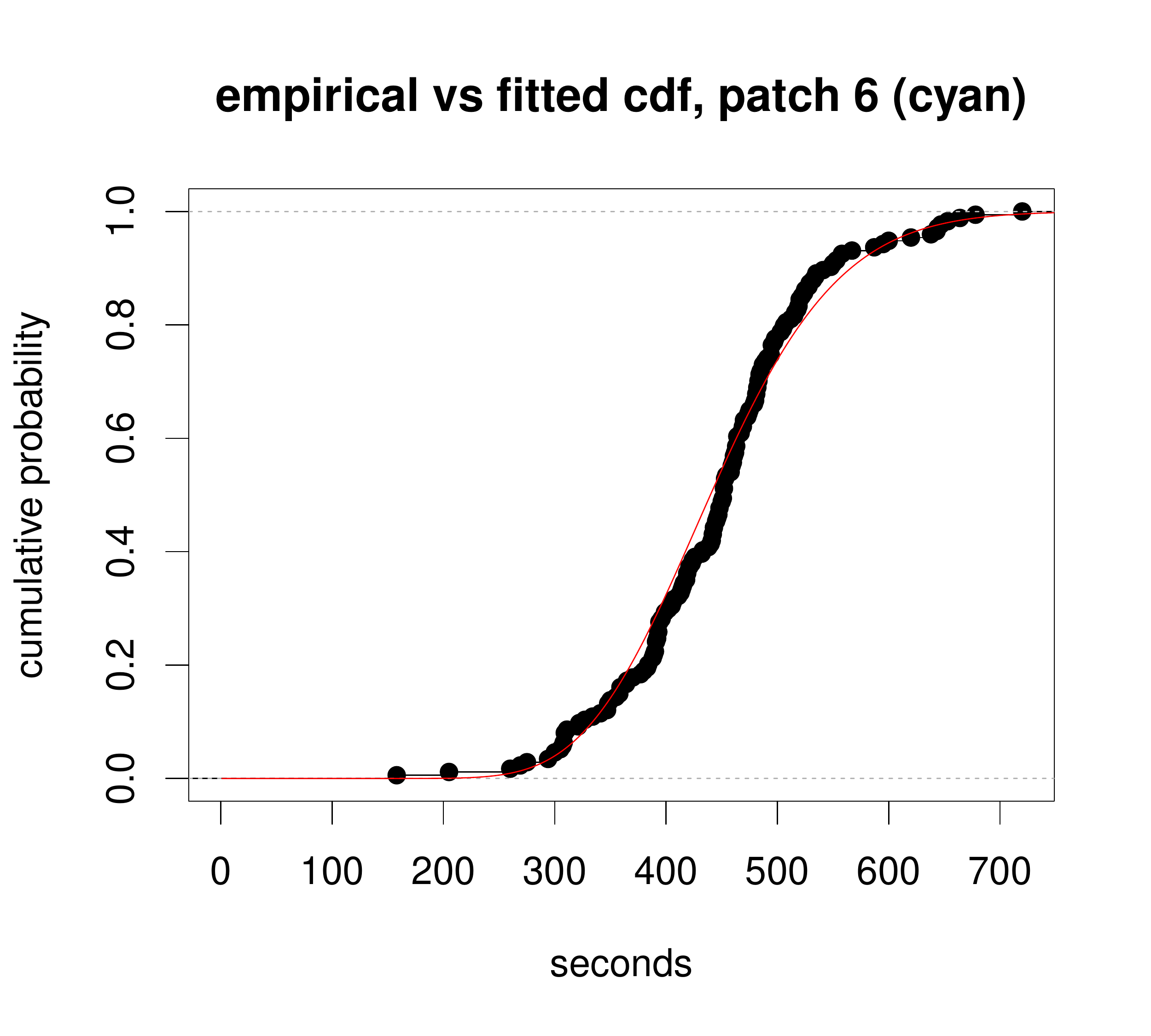} \label{fig: seattle_cdf5}}  \vspace{-0.5cm} \\
			 	\subfloat{\includegraphics[width=\cdfscale]{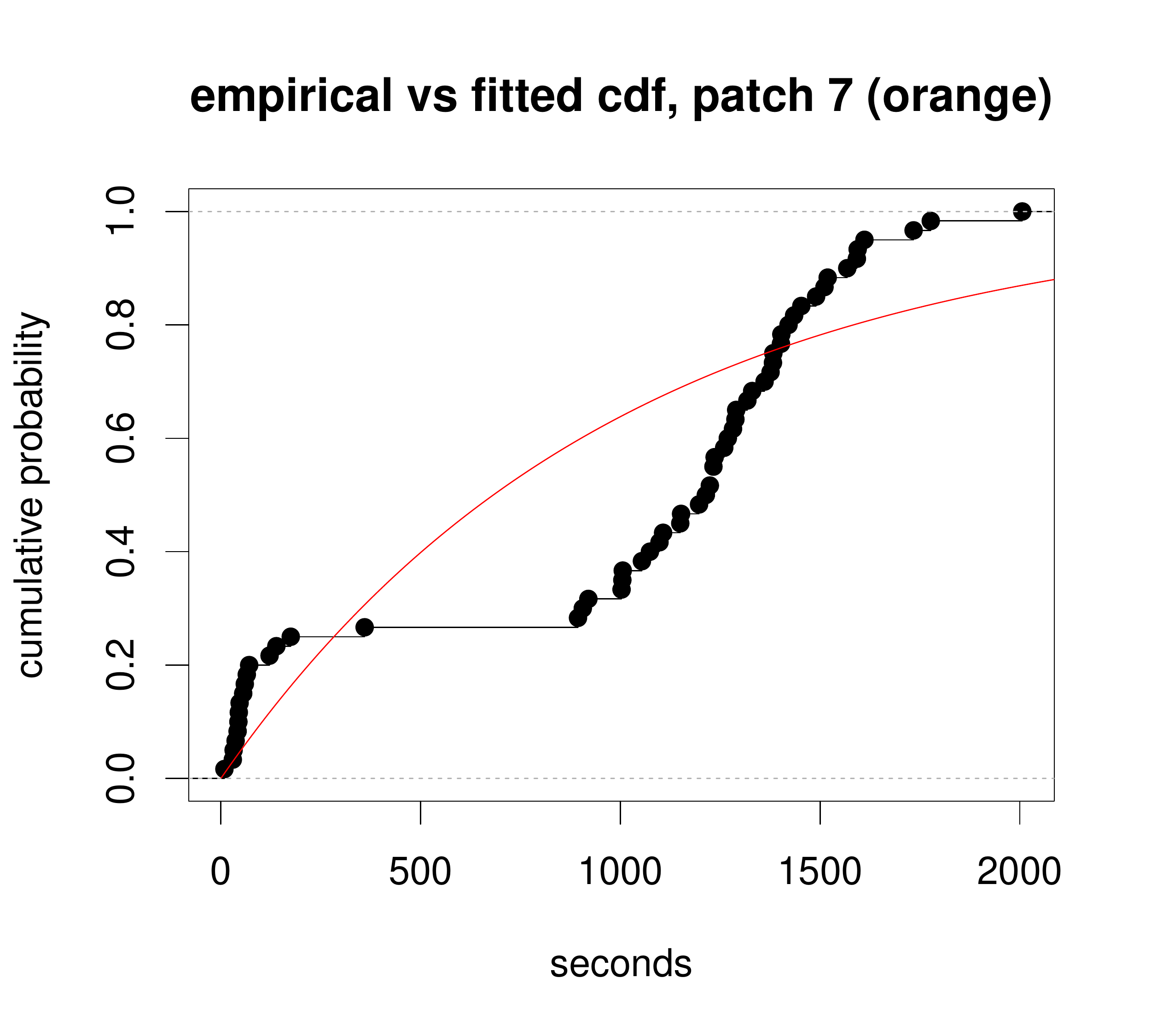} \label{fig: seattle_cdf6}} 
			 & \subfloat{\includegraphics[width=\cdfscale]{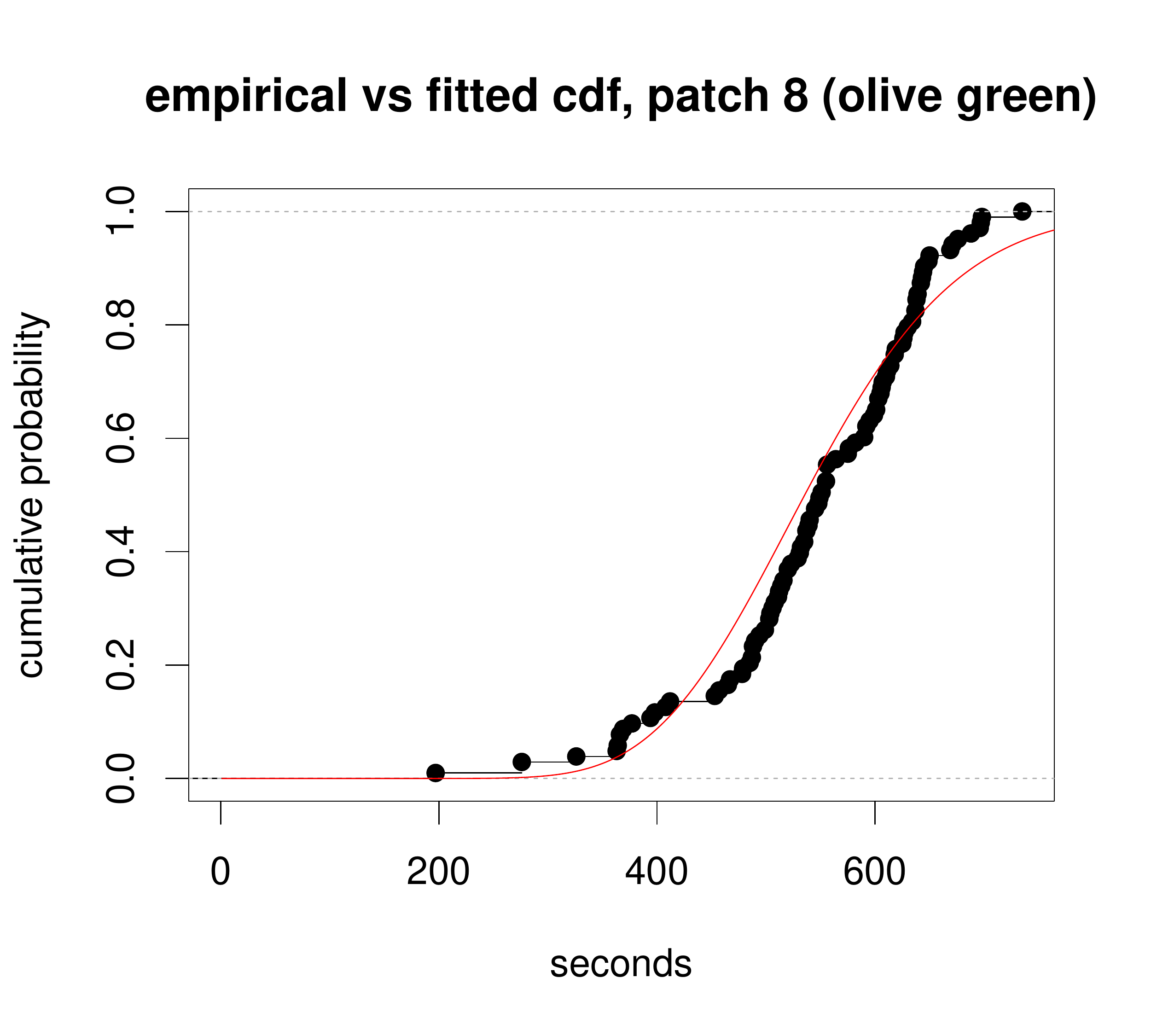} \label{fig: seattle_cdf7}}
			 & \subfloat{\includegraphics[width=\cdfscale]{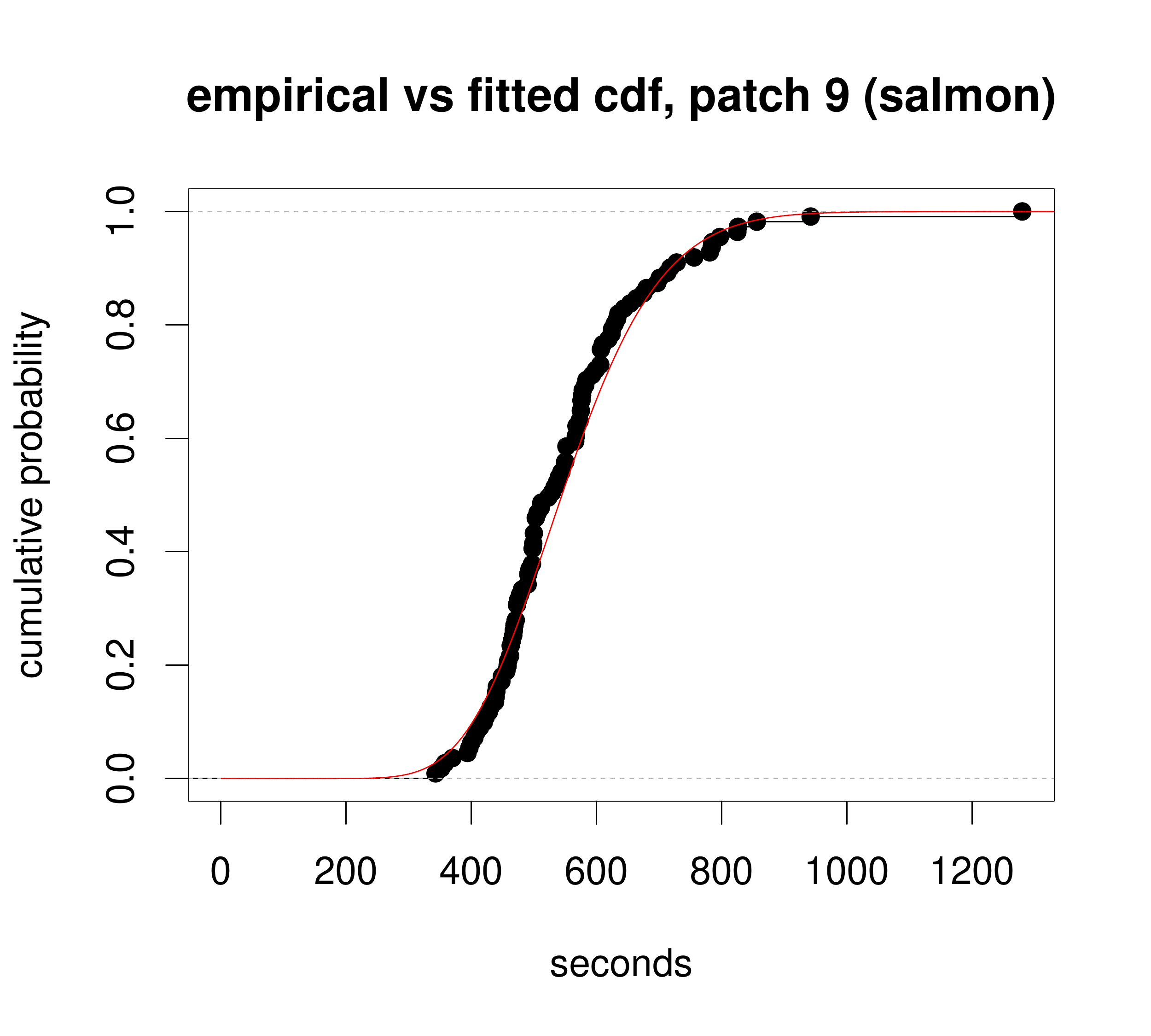} \label{fig: seattle_cdf8}}  \vspace{-0.5cm} \\
			    \subfloat{\includegraphics[width=\cdfscale]{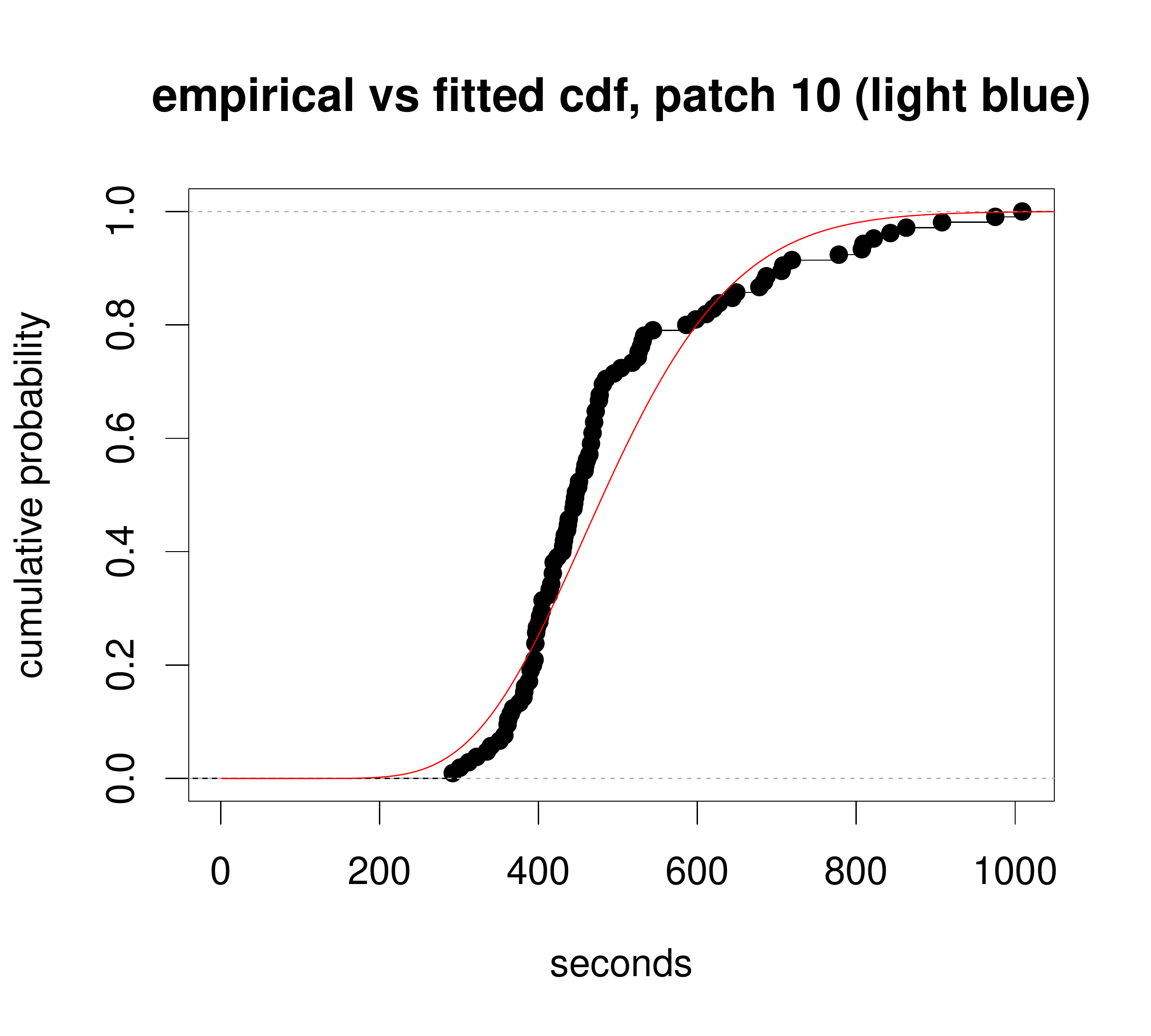} \label{fig: seattle_cdf9}}
			 & \subfloat{\includegraphics[width=\cdfscale]{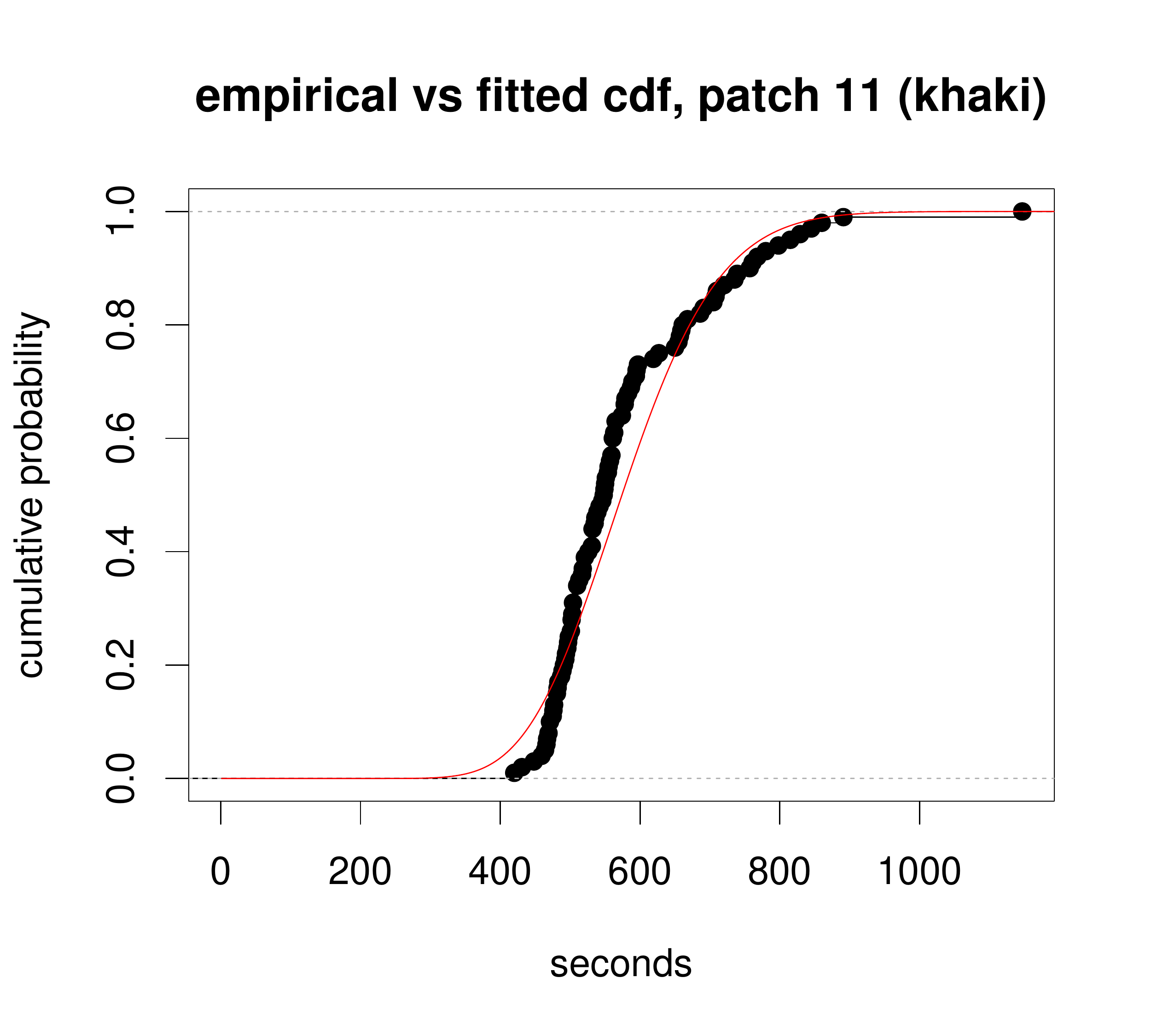} \label{fig: seattle_cdf10}}
			 & \subfloat{\includegraphics[width=\cdfscale]{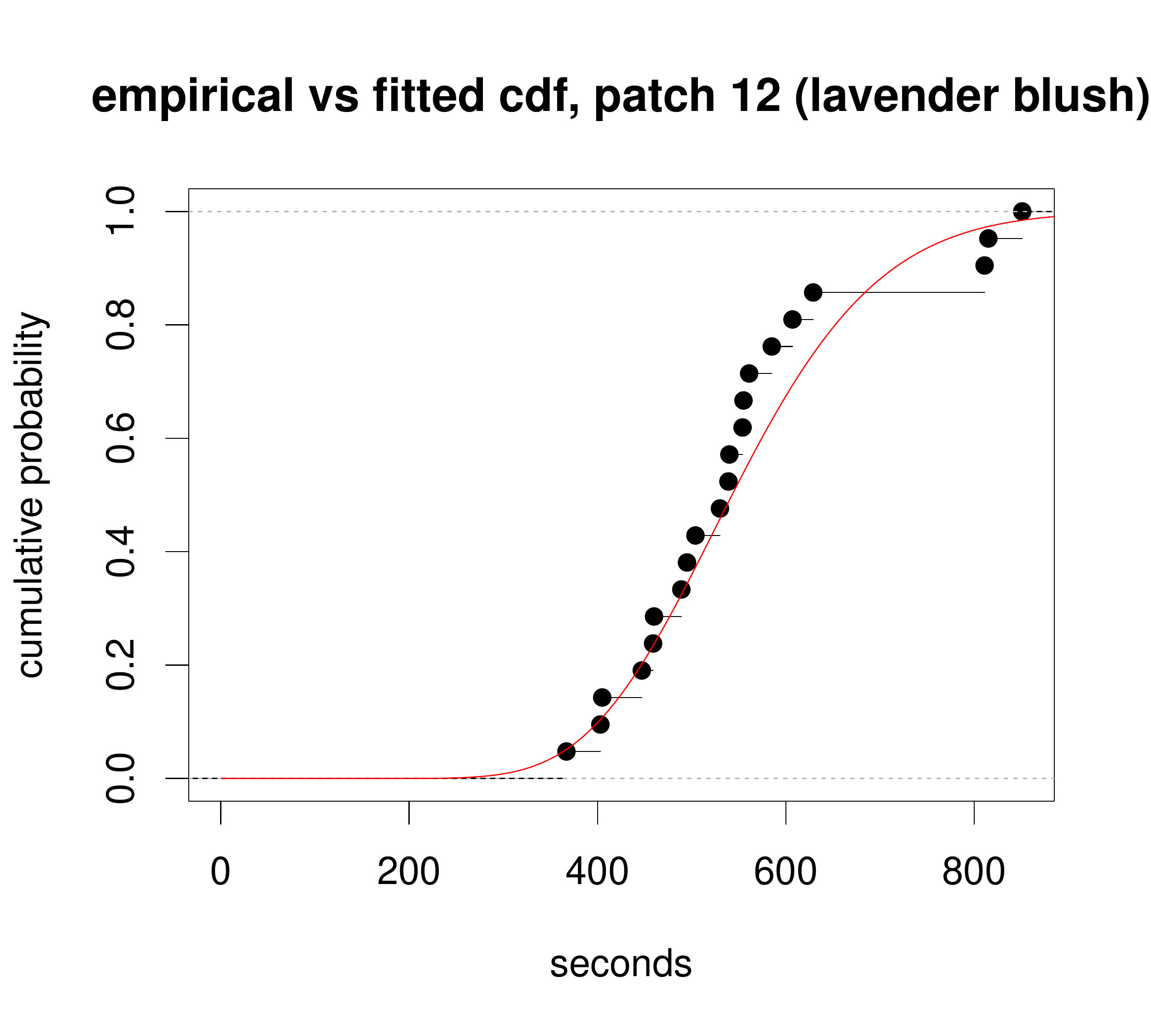} \label{fig: seattle_cdf11}} \vspace{-0.5cm} \\
		
			\end{tabular}
		}
	\end{center}
	\caption{Empirical (black dots) vs fitted (smooth lines) CDF plots for all of the 12 final patches in the Seattle dataset.}
	\label{fig: cdf plots seattle}
\end{figure}
\setlength{\tabcolsep}{6pt} 
\renewcommand{\arraystretch}{1} 

\end{document}